\documentclass[twocolumn]{aastex7}

\usepackage[T1]{fontenc}

\usepackage{bm}        
\usepackage{amssymb, amsmath}   
\usepackage{cancel}
\usepackage{color}

\newcommand\aastex{AAS\TeX}

\DeclareMathAlphabet{\mathsfit}{\encodingdefault}{\sfdefault}{m}{sl}
\SetMathAlphabet{\mathsfit}{bold}{\encodingdefault}{\sfdefault}{bx}{sl}

\newcommand{\vect}[1]{\bm{#1}}

\received{****}
\revised{****}
\accepted{****}
\submitjournal{ApJ}

\shorttitle{\aastex\ Electron acceleration in 3D flare regions}
\shortauthors{Li et al.}

\begin{document}

\title{Energy Conversion and Electron Acceleration and Transport in 3D Simulations of Solar Flares}

\correspondingauthor{Xiaocan Li}
\email{xiaocanli@lanl.gov}

\author[0000-0001-5278-8029]{Xiaocan Li}
\affil{Los Alamos National Laboratory, Los Alamos, NM 87545, USA}
\email{xiaocanli@lanl.gov}

\author[0000-0002-9258-4490]{Chengcai Shen}
\affiliation{Harvard-Smithsonian Center for Astrophysics, 60, Garden Street, Cambridge, MA, 02138, USA}
\email{chengcaishen@cfa.harvard.edu}

\author[0009-0007-0582-7807]{Xiaoyan Xie}
\affiliation{Harvard-Smithsonian Center for Astrophysics, 60, Garden Street, Cambridge, MA, 02138, USA}
\email{xiaoyan.xie@cfa.harvard.edu}

\author[0000-0003-4315-3755]{Fan Guo}
\affiliation{Los Alamos National Laboratory, Los Alamos, NM 87545, USA}
\email{guofan@lanl.gov}

\author[0000-0002-0660-3350]{Bin Chen}
\affiliation{Center for Solar-Terrestrial Research, New Jersey Institute of Technology, 3L King Jr. Blvd., Newark, NJ 07102-1982, USA}
\email{bin.chen@njit.edu}

\author{Ivan Oparin}
\affiliation{Center for Solar-Terrestrial Research, New Jersey Institute of Technology, 3L King Jr. Blvd., Newark, NJ 07102-1982, USA}
\email{ido4@njit.edu}

\author[0000-0002-6628-6211]{Yuqian Wei}
\affiliation{Center for Solar-Terrestrial Research, New Jersey Institute of Technology, 3L King Jr. Blvd., Newark, NJ 07102-1982, USA}
\email{yw633@njit.edu}

\author[0000-0003-2872-2614]{Sijie Yu}
\affiliation{Center for Solar-Terrestrial Research, New Jersey Institute of Technology, 3L King Jr. Blvd., Newark, NJ 07102-1982, USA}
\email{sijie.yu@njit.edu}

\author[0000-0002-5550-8667]{Jeongbhin Seo}
\affiliation{Los Alamos National Laboratory, Los Alamos, NM 87545, USA}
\email{jseo@lanl.gov}

\begin{abstract}
Recent observations and simulations indicate that solar flares undergo extremely complex three-dimensional (3D) evolution, making 3D particle transport models essential for understanding electron acceleration and interpreting flare emissions. In this study, we investigate this problem by solving Parker's transport equation with 3D MHD simulations of solar flares. By examining energy conversion in the 3D system, we evaluate the roles of different acceleration mechanisms, including reconnection current sheet (CS), termination shock (TS), and supra-arcade downflows (SADs). We find that large-amplitude turbulent fluctuations are generated and sustained in the 3D system. The model results demonstrate that a significant number of electrons are accelerated to hundreds of keV and even a few MeV, forming power-law energy spectra. These energetic particles are widely distributed, with concentrations at the TS and in the flare looptop region, consistent with results derived from recent hard X-ray (HXR) and microwave (MW) observations. By selectively turning particle acceleration on or off in specific regions, we find that the CS and SADs effectively accelerate electrons to several hundred keV, while the TS enables further acceleration to MeV. However, no single mechanism can independently account for the significant number of energetic electrons observed. Instead, the mechanisms work synergistically to produce a large population of accelerated electrons. Our model provides spatially and temporally resolved electron distributions in the whole flare region and at the flare footpoints, enabling synthetic HXR and MW emission modeling for comparison with observations. These results offer important insights into electron acceleration and transport in 3D solar flare regions.
\end{abstract}


\section{Introduction}

Solar flares are among the most energetic phenomena within the heliosphere. Within minutes, a large amount of coronal magnetic energy is released and converted into plasma thermal energy and nonthermal particles, heating plasma to high temperature ($>$ 10$^7$ K, \citealp[]{Reeves2011ApJ,Warren2018ApJ}) and producing nonthermal emissions in MW, HXR, and $\gamma$-ray bands~\citep{Lin1976Nonthermal,Krucker2010Measure,Emslie2012ApJ,Aschwanden2016ApJ}. Understanding nonthermal electron acceleration and associated emissions is of central importance in solar flare research. While decades of research have been dedicated to addressing this problem, a comprehensive model, especially in 3D, remains elusive.

Our current understanding of eruptive solar flares has been generally placed in the context of the standard model of eruptive flares~\citep{Carmichael64,Sturrock66,Hirayama74,Kopp76}. In the standard model, magnetic reconnection and associated violent energy release are thought to occur above the flare loops, particularly in an elongated current sheet region driven by the ejection of a large flux rope, which may appear as a coronal mass ejection (CME) observed by white light coronagraphs. An excellent example that strongly supports this model is the SOL2017-09-10 flare~\citep[e.g.,][]{Chen2020ApJ, Chen2020Measurement}. The observations reveal a close resemblance between the observed structures and those in the standard model, including the flare loop, hot reconnecting current sheet region, and the flux rope/CME bubble ejected into the interplanetary space. Additionally, the observations suggested that a large number of electrons~\citep{Fleishman2022Natur} and ions~\citep{Omodei2018ApJ,Kocharov2020ApJ} are accelerated to nonthermal energies. The nonthermal electrons produce HXR and MW emissions observed by \textit{RHESSI} and \textit{EOVSA}~\citep{Chen2020ApJ,Chen2020Measurement,Yu2020ApJ}. These HXR/MW-emitting electrons tend to be concentrated at the flare looptop region~\citep{Chen2020Measurement,Fleishman2022Natur,Chen2024Energetic}, suggesting either efficient acceleration and/or effective trapping of energetic electrons therein. Nonthermal electrons responsible for these emissions tend to develop power-law or double power-law energy spectra~\citep{Chen2021Energetic}. These features are not specific to the SOL2017-09-10 flare event. Instead, modern spectroscopic observations have suggested these features are common among solar flares and are key to measuring the success of the relevant numerical simulations~\citep[see, e.g.,][for a review]{Gary2023New}.

Based on the standard model and emission observations, many mechanisms have been proposed to explain particle acceleration in solar flares, including magnetic reconnection~\citep{Drake2013Power,Li2018Large,Li2022Modeling,Seo2024Proton}, reconnection-driven turbulence~\citep{Petrosian2012Stochastic,Fleishman2020Sci}, termination shocks driven by reconnection outflows~\citep{Guo2012Particle,Chen2015Particle,Kong2019,Kong2020ApJ}, and large-scale plasma waves in the flare loops~\citep{Fletcher2008ApJ,Yu2019Possible}. All these mechanisms could play a role in accelerating nonthermal electrons. Which mechanism dominates electron acceleration is still not well understood. Moreover, each of these mechanisms is a complex topic on its own, making it even more challenging to study them together in solar flares.

Another challenge in studying electron acceleration in solar flares lies in the large-scale separation~\citep{Li2021PoP}. Processes like magnetic reconnection and turbulence dissipation occur at kinetic scales, typically smaller than 1 meter, whereas solar flares can be over 100 Mm. The eight-order spatial (and temporal) scale separation forbids any individual simulation from fully addressing the nonthermal electron acceleration during solar flares. For example, kinetic simulations (either fully kinetic or hybrid) can self-consistently study particle acceleration but are limited to systems up to $\sim 10^3$ ion inertial lengths~\citep[$\sim 1$ km,][]{Li2019Formation,Zhang2024PhRvL}. In contrast, although MHD simulations can capture the large-scale flare geometries and dynamics, the finest grids of MHD simulations of solar flares have grid cells of tens of km~\citep[e.g.,][]{Cheung2019NatAs}, and they cannot study nonthermal particle acceleration and associated emissions. To better interpret nonthermal emissions from solar flares, macroscopic energetic-particle models are needed to capture particle acceleration and transport within a realistic, time-evolving MHD flare geometry. These macroscopic particle models should incorporate acceleration and transport processes learned from self-consistent kinetic simulations~\citep[e.g.,][]{Drake2006Electron,Guo2014Formation,Dahlin2014Mechanisms,Dahlin2017Role,Li2015Nonthermally,Li2017Particle,Li2018Roles,Li2019Formation}.

Several such macroscopic models have been proposed for studying particle acceleration in a large-scale reconnection layer and solar flares. One approach is based on equations derived from particle guiding-center drift motions, and it has successfully explained the enhancements of energetic particle fluxes associated with small-scale flux ropes in the solar wind~\citep{LeRoux2015Kinetic,LeRoux2016Combining,LeRoux2018Self}. Recent progress has demonstrated that it is possible to couple this model with MHD equations, enabling feedback from energetic particles to the MHD flow~\citep{Drake2019, Arnold2019,Yin2024Simultaneous}. The resulting \texttt{kglobal} model is capable of reproducing extended power-law energy spectra in solar flares reconnection region and heliospheric current sheet~\citep{Arnold2021PRL,Yin2024Simultaneous}. Another approach relies on the conservation of the first and second adiabatic invariants~\citep{Drake2013Power,Zank2014Particle,Zank2015Diffusive}. The resulting transport equations for energetic particles are capable of evolving the anisotropic particle distributions and explaining the local enhancements of energetic particle fluxes in a ``sea'' of magnetic islands or flux ropes in the solar wind~\citep{Zhao2018Unusual,Zhao2019Particle,Adhikari2019Role}. If pitch-angle scattering is strong enough to make the particle distributions nearly isotropic, Parker-type transport equations can be derived using this approach~\citep{Parker1965Passage,Zank2014Particle}. In Parker's transport equation, the first-order acceleration is due to flow compression. Solving these transport equations with MHD simulations has given results successfully explaining particle acceleration and transport in large-scale reconnection layers~\citep{Li2018Large,Seo2024Proton}, flare termination shocks~\citep{Kong2019,Kong2020ApJ,Kong2022Model}, and the entire flare region~\citep{Li2022Modeling}. In such an environment, nearly isotropic nonthermal particle distributions can be assumed due to the frequent scattering by magnetic turbulence~\citep{Li2019Formation}. The results show a general agreement with the observed electron spectra, spatial distributions, and emission signatures in the coronal region and flare footpoints.

However, it is important to realize that coronal reconnection and nonthermal emission can involve significant 3D effects. Both observations~\citep{Kontar2017,Cheng2018ApJ,Warren2018ApJ,French2019} and numerical simulations~\citep{Daughton2011Role,Guo2015Particle,Dahlin2015Electron,Huang2016Turbulent,Li2019Formation,Zhang2021Efficient,Wang2023Three,Wang2025Basic} suggest that flare reconnection spontaneously develops turbulence and undergo 3D evolution~\citep[e.g.,][]{Titov99}. New physical processes relevant to flare energy release and particle acceleration could also emerge during 3D flare evolution. For example, the supra-arcade downflows (SADs) often observed in solar flares~\citep{McKenzie2000,Savage2011Re,Xie2022ApJ} from a face-on viewing perspective and emerge only in 3D MHD simulations~\citep{Shen2022}. These 3D dynamics often leave imprints on the resulting nonthermal emissions. The HXR emissions at the flare footpoints are often not ``ribbon-like'', as observed in other wavebands. Even for a few rare HXR ribbons~\citep{Liu07Eruption,Krucker11High}, the emission is not uniform, indicating the importance of considering electron acceleration and transport during 3D reconnection. Additionally, joint observations using Solar Orbiter/STIX, EOVSA, and/or the Hard X-ray Imager (HXI; \citealt{ZhangZ2019}) onboard the Advanced Space-based Solar Observatory (ASO-S; \citealt{Gan2019}) from different viewing perspectives and emission bands are now possible~\citep{Mondal2024Joint,Ryan2024}, highlighting the need for further modeling efforts to explain the observations and better understand energy release and electron acceleration. Thus, it is necessary to build 3D models of nonthermal electron acceleration and transport, as well as associated nonthermal emissions during solar flares.

In this paper, we solve Parker's transport equation using temporally evolving background magnetic fields and plasma flows obtained in an MHD simulation of solar flares to investigate electron acceleration and transport processes in 3D flare regions. We adopt the same 3D MHD simulation that successfully explained the origin of the SADs in the flare region~\citep{Shen2022}. The 3D MHD simulation is particularly valuable for particle modeling because it reveals the most important regions for energetic particle acceleration, including the reconnecting current sheet, termination shock, and turbulent flare looptop, which contains SADs. In Section~\ref{sec:num}, we describe the setup for the MHD simulations and the energetic particle transport modeling. In Section~\ref{sec:results}, we present our simulation results. We have examined the energy conversion, flow compression distribution, and plasma turbulence in the 3D system. Additionally, we have examined the electron energy spectra, electron spatial distribution maps, and precipitated electrons at the flare footpoints, as well as primary electron acceleration mechanisms. In Section~\ref{sec:con}, we discuss the model setup, model results, and possible future works.

\section{Numerical Simulations}
\label{sec:num}

We adopted the same MHD simulation setup as in Case A in~\citet{Shen2022}. The simulation is initialized with a vertical Harris-type current sheet in 2.5D (along the $y-$direction on the $x-y$ plane; the $z$ direction is homogeneous). After an initial perturbation, magnetic reconnection occurs in the current sheet, resulting in the formation of the classic Kopp-Pneuman configuration, which includes the two-ribbon flare loops at the bottom and an upwards-extending reconnection current sheet above. We then use the magnetic configuration of the 2.5D simulation results as the initial condition to build the 3D model by symmetrically extending the primary variables along the 3rd, $z$-direction. The 3D system then self-consistently evolves, revealing various dynamic features of plasma in solar flares. The MHD simulations are performed using the Athena code~\citep{Stone2008Athena} and include Ohmic dissipation, thermal conduction, radiative cooling, coronal heating terms, and static gravity in the source term of the MHD energy equation. Interested readers are referred to~\citet{Shen2022} for details on the numerical algorithms. The magnetic field is line-tied to the photosphere at the bottom boundary, and the plasma does not slip either. The other boundaries are all open, allowing the plasma and magnetic flux to enter or exit freely.

The simulations are performed on a uniform Cartesian grid with domain $x\in[-0.5, 0.5]$, $y\in[0, 1]$, and $z\in[-0.25, 0.25]$ for the 3D simulation. It has $576 \times 576 \times 288$ cells. We normalize the simulations by setting the characteristic length $L_0=150$ Mm, magnetic field strength $B_0=9.9$ Gauss, and number density of plasma $n_0=2.5\times10^{8}\text{ cm}^{-3}$. The resulting grid resolution is $\sim 0.3$ Mm. Plasma velocity is normalized with the characteristic Alfv\'en speed $V_{A0}=B_0/\sqrt{\mu_0n_0m_p}=1365.3$ km/s, where $m_p$ is the proton mass. The evolution time is normalized with $T_0=L_0/V_{A0}=109.86$ s, and the entire 3D simulation represents the system evolution for about 690 s (6.3 $T_0$). Since the initial condition of the 3D simulation is a snapshot of the 2.5D simulation lasting $17T_0$, the total simulation time $23.3T_0$. We carry out the simulations with a uniform resistivity $\eta=2\times10^{-5}$, resulting in a magnetic Reynolds number (Lundquist number) $R_m=5\times10^4$. We have also performed particle modeling using an MHD simulation with $R_m=10^4$ (case B in~\citet{Shen2022}) and found the acceleration to be much weaker. Thus, the results for $R_m=10^4$ will not be presented here. We collect the MHD fields every $0.02T_0$ and use them to solve Parker's transport equation,
\begin{equation}
  \frac{\partial f}{\partial t} + (\vect{V}+\vect{V}_d)\cdot\nabla f
  - \frac{1}{3}\nabla\cdot\vect{V}\frac{\partial f}{\partial\ln p}
  = \nabla\cdot(\vect{\kappa}\nabla f) + Q,\label{equ:parker}
\end{equation}
where $f(\vect{x}, p, t)$ is the particle distribution function, which is a function of particle position $\vect{x}$, momentum $p$, and time $t$; $\vect{V}$ is the bulk plasma velocity (provided by the 3D MHD simulation); $\vect{V}_d$ is the particle drift velocity, which includes $\nabla B$ drift, curvature drift, and parallel drift~\footnote{$V_d$ is $\sim 0.03$ km/s for 10 keV electrons and a few km/s for MeV electrons in our simulations. It is much smaller than the Alfv\'en speed $V_{A0}$ and thus not particularly important for particle evolution.}; $\vect{\kappa}$ is the spatial diffusion coefficient tensor; $Q$ is the particle source term. We solve Parker's transport equation without evolving the particle pitch angles because earlier fully kinetic simulations have demonstrated that self-generated turbulence in 3D reconnection~\citep{Daughton2011Role,Guo2015Particle,Dahlin2015Electron,Huang2016Turbulent,Li2019Formation,Zhang2021Efficient} produces strong pitch-angle scattering and weak anisotropy for high-energy electrons~\citep{Li2018Roles,Li2019Formation}.

Depending on the local magnetic field direction $\vect{b}$, the spatial diffusion is determined by the diffusion coefficient tensor $\kappa_{ij} = \kappa_\perp\delta_{ij} - (\kappa_\perp-\kappa_\parallel)b_ib_j$, where $\kappa_\parallel$ and $\kappa_\perp$ are the diffusion coefficients along and across the magnetic field lines, respectively. According to the quasi-linear theory~\citep{Jokipii1971Propagation,Giacalone1999Transport}, $\kappa_\parallel\approx1.622v^{4/3}L_c^{2/3}/(\Omega_0^{1/3}\sigma^2)$ when magnetic turbulence is fully developed and has an isotropic power spectrum $\sim k^{-5/3}$, where $v$ is the particle speed, $\Omega_0=eB_0/(m_ec)$ is the particle gyrofrequency, $L_c$ is the turbulence correlation length, and $\sigma^2=\left<\delta B^2\right>/B_0^2$ is the turbulence amplitude. In our simulations, we set $\kappa_\parallel=\kappa_{\parallel 0}(v/v_0)^{4/3}(B_0/B)^{1/3}$, where $v_0$ is the speed of the initial mono-energetic particles (10 keV, see Section~\ref{subsec:acc} for more discussions). $\kappa_{\parallel 0}$ is the corresponding parallel diffusion coefficient, and we assume $L_c\approx1000$ km and $\sigma^2=1$ (see discussions in Section~\ref{subsec:turb}). Normalized by $\kappa_0=L_0V_\text{A0}\approx2.048\times10^{14}\text{ m}^2\text{ s}^{-1}$, $\tilde{\kappa}_{\parallel0}=\kappa_{\parallel0}/\kappa_0=3.28\times10^{-3}$ in the simulations. We set $\kappa_\perp/\kappa_\parallel=0.01$, as test-particle simulations have suggested that particle parallel diffusion is much faster than cross-field diffusion~\citep{Giacalone1999Transport}.

To solve Parker's transport equation, we evolve a large ensemble of pseudo-particle electrons using stochastic differential equations (SDEs) that are formally equivalent to the Fokker-Planck form of Parker's transport equation. This numerical method has been documented extensively in the literature~\citep{Zhang1999Markov,Florinski2009Four,Pei2010General,Guo2010,Li2018Large,Li2022Modeling,Kong2019,Kong2020ApJ,Seo2024Proton}. The particles are called pseudo particles because they sample the distribution function $fp^2$ instead of $f$. We refer interested readers to~\citet{GPAT_GitHub} for implementation details on solving the SDEs. We assume that injected particles originate from the thermal plasma and are subsequently accelerated. Capturing these injection mechanisms requires self-consistent, fully kinetic simulations, which have shown that electrons enter the system either by streaming along the exhaust boundaries or by encountering strong ideal or non-ideal electric fields near the reconnection X-points~\citep{Egedal2013Review,Guo2019Determining,Zhang2021Efficient}. To mimic these injection processes, we continuously inject particles in the flare current sheet (the source $Q$), where the current density is the strongest. We also avoid injecting the particles in the flare loops. Specifically, we inject particles in the regions where the current density $|\vect{j}|>100$ and $y>0.25$ (or $>37.5$ Mm in physical units) during particle modeling. We set a constant $Q=1.835\times 10^7$ pseudo particles/s in our simulations. Keep in mind that particle injection process can be more complex. For example, it likely depends on plasma parameters, such as plasma $\beta$ and guide field, and injected particle fluxes should scale with the time-varying reconnection rate. Addressing these dependencies is a complex issue that warrants additional kinetic studies. During the simulations, we turn off particle acceleration when $y<0.25$ (or $<37.5$ Mm in physical units) to avoid electron acceleration in the dense flare loops, where radiation cooling and collisional energy loss could prevent efficient electron acceleration. Note that during particle modeling, we use a particle-splitting technique to improve the statistics at high energies. When a particle's momentum reaches $2\times1.2^{n_\text{split}}$ times its initial value $p_0$, it is split into two identical particles with half the original weight before splitting, where $n_\text{split}$ is the number of times the particle has been split. These two particles will evolve independently due to the stochastic nature of the transport process. In total, about 12.7 billion pseudo particles are continuously injected into the system. Due to particle escape from the boundaries and the particle-splitting technique, approximately 6.8 billion pseudo particles with various weights remain in the system at the end of the simulations.

\section{Simulation Results}
\label{sec:results}

\subsection{Overview of the 3D MHD simulation}
\label{subsec:overview}

\begin{figure*}[ht!]
  \centering
  \includegraphics[width=\linewidth]{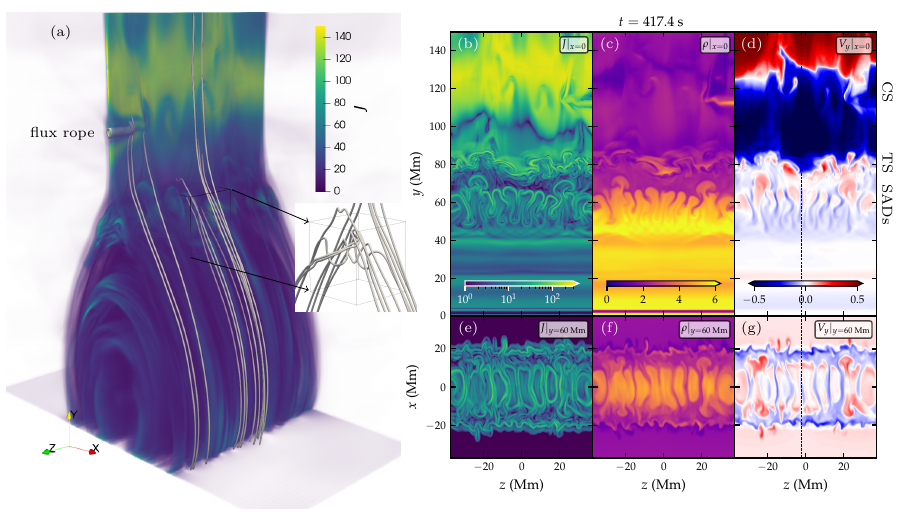}
  \caption{\label{fig:mhd}
    Overview of the 3D MHD simulation results~\citep{Shen2022}. (a) 3D rendering of the current density in the flare region. The white lines indicate sample magnetic field lines. A flux rope is generated during flare reconnection. The embedded plot shows the twisted magnetic field lines near the looptop region. (b)--(d) 2D slices of the current density $J$, plasma density $\rho$, and plasma velocity $V_y$ in the $y-z$ plane at $x=0$. (e)--(g) 2D slices of $J$, $\rho$, and $V_y$ in the $z-x$ plane through the supra-arcade downflows (SADs) regions at $y=60$ Mm. The texts on the right indicate reconnecting current sheet (CS), termination shock (TS), and SADs regions. The vertical dashed lines in (d) \& (g) indicate the $z$-cuts shown in Figure~\ref{fig:econv_mhd_xy}.
  }
\end{figure*}

Figure~\ref{fig:mhd} shows the overview of the 3D MHD simulation results. The 3D rendering of the current density $J$ shows the typical flare geometry (Figure~\ref{fig:mhd} (a)), including the reconnecting current sheet (CS) and the flare arcades. $J$ intensifies in the CS, leading to fast reconnection and the onset of tearing instability, manifested by the formation of the flux rope. Zooming into the above-the-looptop region, we find twisted magnetic field lines, indicating the generation of plasma turbulence. Previous studies have demonstrated that such turbulence appears as a natural result of the development of various MHD instabilities \citep{Shen2022,Ruan2023,Shibata2023}. The right panels of Figure~\ref{fig:mhd} show 2D slices through the flare region. Figures~\ref{fig:mhd} (b)--(d) show 2D slices through the center of the flare region ($x=0$). Besides the CS, $J|_{x=0}$ also intensifies and shows clear structures near the termination shock (TS, $70\leq y\leq90$ Mm) and the finger-like bubbles (representing SADs) (Figure~\ref{fig:mhd} (b)). The density slice $\rho|_{x=0}$ shows the high-density flare loops, which contrast with the low densities in the rest of the flare region. The most interesting region is the interface in between, where the SADs are self-generated due to a mixture of Rayleigh-Taylor instability (RTI) and the Richtmyer-Meshkov instability (RMI)~\citep{Shen2022}. The $V_y|_{x=0}$ slice shows the bidirectional reconnection outflows, the sharp termination of the downward outflows near the TS, and the SADs and accompanying upflows sandwiched between the SADs. Note that the downflows generated by reconnection are not uniform due to 3D effects. They are relatively weak in some regions (e.g., around $z=-20$ Mm), which will affect particle transport, and will be discussed in Section~\ref{subsubsec:dist}. Additionally, $V_y=0$ in the CS region shows the primary X-line wandering around, highlighting a distinct feature caused by 3D effects. Figures~\ref{fig:mhd} (e)--(g) show the 2D slices through the center of the SADs region ($y=60$ Mm), where the SADs are manifested by the bubble-like structures. It is noteworthy that the current density tends to be stronger at the boundary of the structures. The low-density regions are sandwiched between high-density regions (Figure~\ref{fig:mhd} (f)). Comparing this with panel (g), we find that the SADs ($V_y<0$) primarily bring the low-density plasmas to the lower flare loops, while the accompanying upflows bring the dense plasmas upwards. The simultaneous existence of upflows and downflows in the interface where SADs habitat has been reported in the observations of \citet{Xie2025ApJ}. We expect that the upward and downward flows will help transport energetic particles in the above-the-looptop region. Additionally, the density jumps with accompanying velocity jumps between different regions indicates that the plasma could be compressed at the boundary layers, where particles can be energized due to flow compression~\citep{Li2018Large,Li2022Modeling,Seo2024Proton}.

\subsection{Energy Conversions in the 3D flare region}
\label{subsec:energy}

Since flow compression drives both particle acceleration in Parker's transport equation (Eq.~\ref{equ:parker}) and plasma heating in the MHD description~\citep[e.g.,][]{Birn2009Energy}, these two processes are coupled. Therefore, examining plasma heating in the 3D MHD simulation can help pinpoint where the acceleration is likely to occur. As the ultimate energy source of solar flares is the magnetic energy stored in the solar corona, it is essential to analyze how the magnetic energy is converted into plasma heating. To study the energy conversion processes, we use the energy equations for the magnetic field, bulk kinetic energy $K$, and internal energy $U$.
\begin{align}
  \frac{\partial}{\partial t}\left(\frac{B^2}{2}\right) & = -\nabla\cdot(\vect{E}\times\vect{B})
  - \vect{V}\cdot(\vect{\vect{j}\times\vect{B}}) - \eta j^2, \label{equ:eneb}\\
  \frac{\partial K}{\partial t} & = -\nabla\cdot(K\vect{V}) +
  \vect{V}\cdot(\vect{\vect{j}\times\vect{B}} - \nabla P),  \label{equ:enek}\\
  \frac{\partial U}{\partial t} & = -\nabla\cdot\left[(U + P)\vect{V}\right] +
  \vect{V}\cdot\nabla P + \eta j^2,  \label{equ:eneu}
\end{align}
where $E$ is the electric field, $\eta$ is the resistivity, and $P$ is the pressure. Similar methods have been used to study energy conversion and plasma heating in 3D magnetic reconnection during solar eruptions~\citep{Birn2009Energy,Reeves2019Exploring}. Note that the MHD simulation also includes other energy source terms, such as thermal conduction, radiative cooling, and coronal heating~\citep{Shen2022}, which do not contribute to the energy conversion between different energies in the MHD simulation and are thus ignored in the discussion here. For example, heating conduction is primarily responsible for redistributing the thermal energy in different regions~\citep{Reeves2019Exploring}.


Eq.~\ref{equ:eneb} shows that the magnetic energy is converted into Poynting flux $\vect{E}\times\vect{B}$, bulk kinetic energy through $\vect{V}\cdot(\vect{\vect{j}\times\vect{B}})$, and internal energy through $\eta j^2$. The Poynting flux can propagate to other locations and be converted directly into magnetic energy or indirectly into bulk kinetic or internal energies. $\vect{V}\cdot(\vect{\vect{j}\times\vect{B}}) =\vect{j}\cdot(-\vect{V}\times\vect{B})$ is the work done by the motional electric field $\vect{E}_m=-\vect{V}\times\vect{B}$. $\eta j^2$ is the work done by the resistive electric field $\eta\vect{j}$. In the simulation setup, although $\eta$ is not as small as in the highly conductive corona, it is still only $2\times10^{-5}$. Thus, this term plays a minimal role in heating the plasmas and will not be discussed further. Part of $\vect{j}\cdot\vect{E}_m$ becomes the bulk flow energy flux and gets carried away by plasma flows. Comparing Eqs.~\ref{equ:enek} and~\ref{equ:eneu}, we find that the bulk kinetic energy can then be converted into the internal energy through the pressure work term $\vect{V}\cdot\nabla P$. However, it does not necessarily result in a net increase in internal energy. Part of $\vect{V}\cdot\nabla P$ is converted into the enthalpy flux $(U+P)\vect{V}$ instead. The net increase of the internal energy is $-\nabla\cdot(P\vect{V}) + \vect{V}\cdot\nabla P=-P\nabla\cdot\vect{V}$, which is due to adiabatic compression. For the discussion below, we will focus on the energy conversion rates $\vect{j}\cdot\vect{E}_m$, $\vect{V}\cdot\nabla P$, and $-P\nabla\cdot\vect{V}$, and leave a more detailed analysis that includes all the terms (e.g., Poynting flux and enthalpy flux) to a future study.

\begin{figure*}[ht!]
  \centering
  \includegraphics[width=0.75\linewidth]{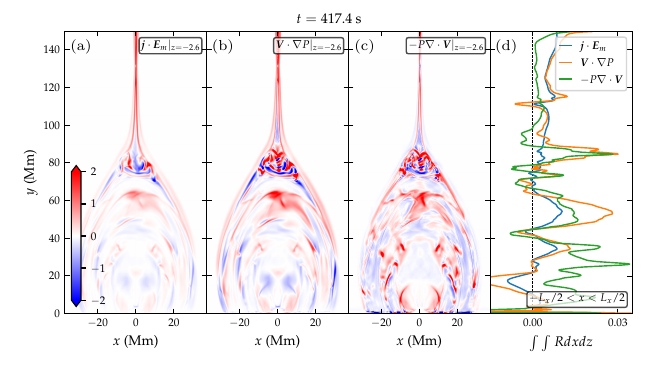}
      \caption{\label{fig:econv_mhd_xy}
  Energy conversion rates in the MHD simulation. (a)--(c) 2D slices of the energy conversion terms at $z=-2.6$ Mm, as indicated in Figure~\ref{fig:mhd} (d) \& (g). (d) The distribution of different terms along the $y$-direction. $R$ in the label along the $x$-axis indicates either of the energy conversion rates. The integral is done over the entire $x$-range.
  }
\end{figure*}

\begin{figure*}[ht!]
  \centering
  \includegraphics[width=0.75\linewidth]{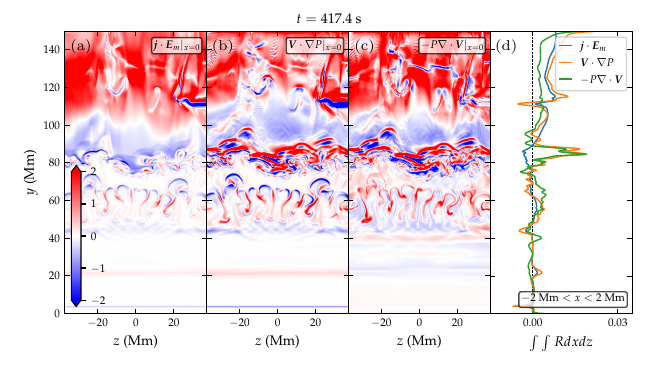}
  \caption{\label{fig:econv_mhd_yz}
  Similar to Figure~\ref{fig:econv_mhd_xy} but from a different perspective. (a)--(c) 2D slices of the energy conversion rates in the $y-z$ plane at $x=0$. (d) Similar to Figure~\ref{fig:econv_mhd_xy} (d) but done in the center of the flare region.
  }
\end{figure*}

Figure~\ref{fig:econv_mhd_xy} shows the rates at $t=417.4$ s. Panels (a)--(c) are 2D slices of different terms at $z=-2.6$ Mm, which is through one of the SADs, and panel (d) shows distributions of different terms along $y$. A positive $\vect{j}\cdot\vect{E}_m$ in the current sheet is expected since magnetic energy is released by magnetic reconnection (panel (a)). Once the reconnection outflow slows down above the flare looptop, the process is reversed, and the magnetic field is compressed, and the magnetic energy is increased due to the negative $\vect{j}\cdot\vect{E}_m$. However, the region ($y=80-100$ Mm) with negative $\vect{j}\cdot\vect{E}_m$ is bounded by the layers with positive $\vect{j}\cdot\vect{E}_m$, which is why the 1D distribution shown in Figure~\ref{fig:econv_mhd_xy} (d) does not show a negative region above the Y point at $y\sim 80$ Mm. The negative $\vect{j}\cdot\vect{E}_m$ becomes more obvious once we focus on the central flare region (i.e., $x\sim 0$) in Figure~\ref{fig:econv_mhd_yz}. Near the SAD region ($y\sim$ 60 Mm), magnetic energy can be dissipated, as indicated by the positive $\vect{j}\cdot\vect{E}_m$. Since the SAD region is broadly distributed, the overall magnetic energy conversion caused by SADs is significant (blue curve near $y=60$ Mm in Figure~\ref{fig:econv_mhd_xy} (d)). However, the local energy conversion rate is not as large as in the reconnection region. Similarly, there is significant magnetic energy conversion in the flare loops below $y=40$ Mm, but the local energy conversion rate is low.

Once the magnetic energy is converted into bulk kinetic energy, it can then be converted into the internal energy or the enthalpy flux through $\vect{V}\cdot\nabla P$. Figure~\ref{fig:econv_mhd_xy} (b) \& (d) show that $\vect{V}\cdot\nabla P$ and $\vect{j}\cdot\vect{E}_m$ are comparable in the reconnection layer. In contrast, the compression energization $-P\nabla\cdot\vect{V}$ is concentrated in a narrower layer (Figure~\ref{fig:econv_mhd_xy} (c)), and it is much lower than the other two terms. This result indicates that the magnetic energy is primarily converted into the enthalpy flux in the reconnection layer. The finite $-P\nabla\cdot\vect{V}$ indicates that the bulk flow energy can be converted into plasma internal energy, thus heating the plasmas in the current sheet. Both $\vect{V}\cdot\nabla P$ and $-P\nabla\cdot\vect{V}$ show complex structures near the TS. These complex structures can also be found in 2D MHD simulations~\citep{Kong2020ApJ}, and they could modulate particle acceleration and transport by the TS. Figure~\ref{fig:econv_mhd_xy} (d) shows that both terms strongly peak near the TS, which is expected since both the bulk kinetic energy and enthalpy flux are converted into the internal energy at the TS. Interestingly, both terms also peak near the interface region (at $y\sim60$\ Mm) where SADs are observed, indicating that plasmas could be heated even in the SADs, consistent with SADs observations~\citep{Xie2023Heating}. The additional peak of $-P\nabla\cdot\vect{V}$ in the flare loops ($y<$ 40 Mm in Figure~\ref{fig:econv_mhd_xy} (d)) is due to the compressive layers or shock waves produced by the chromospheric evaporation, manifested strong transient layers shown in Figure~\ref{fig:econv_mhd_xy} (d) below $y=$ 40 Mm. Particles could potentially be energized in these regions, but the collisional energy loss and radiation cooling could also be significant due to the high plasma density. Since these physics processes are not included in the particle modeling, we also turn off the acceleration in the regions $y<37.5$ Mm.

Figure~\ref{fig:econv_mhd_xy} shows that the manifest energy conversion continuously occurs in the center region of the entire flare loop system, particular the CS and flare looptops. To look into the energy conversion therein, we show in Figure~\ref{fig:econv_mhd_yz} the 2D slices of the energy conversion rates through the center of the flare region and the integration of the rates near the center ($-2$ Mm $<x<$ 2 Mm, $-L_z/2<z<L_z/2$). While similar conclusions can be drawn, Figure~\ref{fig:econv_mhd_yz} shows more structures throughout the flare region than Figure~\ref{fig:econv_mhd_xy}. A distinct feature in the CS is a small-scale flux rope (or magnetic island) at $y\sim$ 110 Mm and $z>20$ Mm. $\vect{j}\cdot\vect{E}_m$ shows a bipolar feature, which is because $\vect{E}_m$ switched direction due to the reversed magnetic field direction at the two ends of the magnetic islands. The broadly distributed negative $\vect{j}\cdot\vect{E}_m$ in the low-end of the CS (Fig.~\ref{fig:econv_mhd_xy}(a)) suggests that bulk flow energy is converted into the magnetic field energy, and the local magnetic field becomes stronger towards the TS. The TS is strongly distorted and breaks into multiple layers. Overall, $\vect{V}\cdot\nabla P$ balances $-P\nabla\cdot\vect{V}$ in both the 2D and 1D distributions, indicating that the enthalpy flux is negligible near the TS. In the SADs region, magnetic energy is converted into plasma energy in the SADs, while the flow energy is converted back to the magnetic energy at the tips of the accompanied upflows. Figure~\ref{fig:econv_mhd_yz} (b) shows that $\vect{V}\cdot\nabla P$ is negative at the tips of the upflows, while Figure~\ref{fig:econv_mhd_yz} (c) shows that the compression energization is positive at both the SADs and the tips of the upflows, which is consistent with observations~\citep{2017ApJ...836...55R,Xie2023Heating}. This result suggests that the enthalpy flux in the upflow regions is converted into internal energy through $-P\nabla\cdot\vect{V}$ and bulk kinetic energy through $\vect{V}\cdot\nabla P$, which is then converted into the magnetic energy through $\vect{j}\cdot\vect{E}_m$. The line plots shown in Figure~\ref{fig:econv_mhd_yz} (d) confirm that in the center of the flare region, most of the energy conversions occur above the flare loops, particularly in the flare current sheet and near the TS.

These energy-conversion results show that, although magnetic reconnection in the flare current sheet initiates the release of magnetic energy, flow compression converts that energy into internal energy throughout the entire flare region. In our 3D MHD simulation, TS is the most efficient site for flow-compression-driven internal-energy increases, followed by CS and SADs. In addition to playing a key role in plasma heating, these regions are the potential regions for particle acceleration due to flow compression, which we will address in the following subsections.

\subsection{Flow Compression in the 3D flare region}
\label{subsec:compression}

\begin{figure}[ht!]
  \centering
  \includegraphics[width=\linewidth]{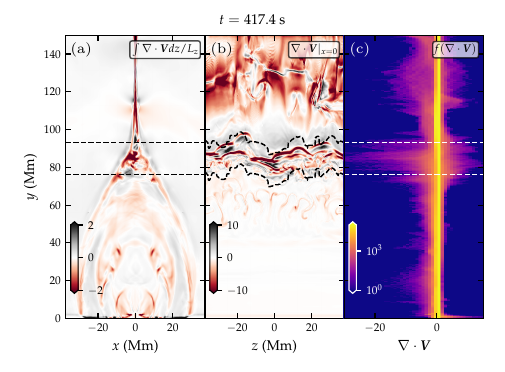}
  \caption{\label{fig:flow_comp}
  Flow compression $\nabla\cdot\vect{V}$ in different regions. (a) $z$-average of $\nabla\cdot\vect{V}$ in the MHD simulation. (b) 2D slices of the flow compression in the $y-z$ plane at $x=0$. (c) Stacked histogram of the flow compression along the $y$-direction. The dashed lines separate the reconnection current sheet, termination shock, and looptop regions (see Section~\ref{subsubsec:mechanisms} for the method). The $y$ locations of the dashed lines in (a) \& (c) are the $z$-average of the dashed lines in (b).
  }
\end{figure}

The discussion above has shown that magnetic energy conversion primarily occurs in the flare current sheet, and the bulk plasma kinetic energy can then be converted into plasma internal energy throughout the flare region through flow compression. Since the only acceleration mechanism to be considered is flow compression in our modeling, its distribution is critical for particle acceleration. Figure~\ref{fig:flow_comp} shows the spatial distributions and stacked histogram of $\nabla\cdot\vect{V}$. Note that panels (a) and (b) have different colorbar ranges. Without the effect of pressure $P$, it is clear that the flow compression is strong in the CS and TS regions. It is weaker near the SADs but broadly distributed. Additionally, the compressive waves in the flare loops can produce layers with strong compression. Figure~\ref{fig:flow_comp} (c) clearly shows $\nabla\cdot\vect{V}$ in the CS can reach $-10T_0^{-1}$, while it can be over $-20T_0^{-1}$ at the TS, where $T_0=109.9$ s is the time normalization of the MHD simulation. A value of $-10T_0^{-1}$ means that particle momentum can increase e-fold every 33 seconds if they stay in the same region and get energized following $dp/dt=-(p/3)\nabla\cdot\vect{V}$. In reality, particles cannot be accelerated so efficiently because they will be constantly scattered and cannot stay in the same region for a long time. For shock acceleration, particles can only get one acceleration during one crossing of the shock. Most of the time, particles are not being accelerated. Anyhow, the large amplitude of $\nabla\cdot\vect{V}$ shows the potential of the 3D region to accelerate electrons to high energies if they can be trapped in local regions, e.g., by plasma turbulence or magnetic structures like plasmoids.

\subsection{Turbulence in the flare region}
\label{subsec:turb}

\begin{figure*}[ht!]
  \centering
  \includegraphics[width=0.9\linewidth]{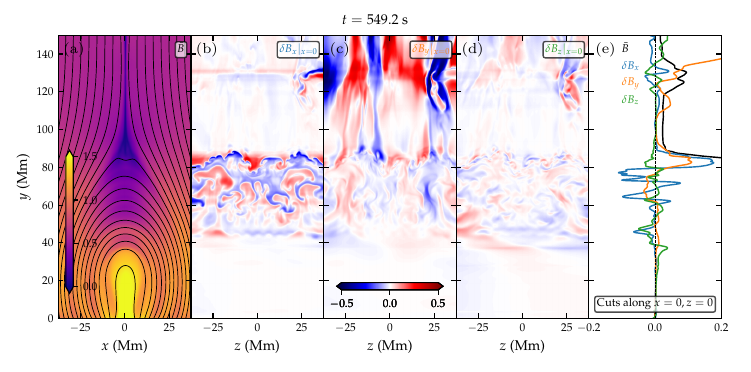}
  \caption{\label{fig:turb}
  Turbulent magnetic fluctuations in the flare region. (a) The magnitude of the mean magnetic field by averaging along the $z$-direction. The magnetic field lines are traced using the mean field $\bar{\vect{B}}$. (b)--(d) 2D slices of the magnetic field fluctuations $\delta B$ in the $y-z$ plane at $x=0$. (e) 1D cut of the mean magnetic field and the turbulent magnetic field components along $x=0$ and $z=0$. The black line is cut off at $y<85$ Mm because $\bar{B}$ is much larger than the turbulent magnetic fluctuations.
  }
\end{figure*}

To examine the turbulence in the 3D flare region, we first average the magnetic field along the $z$-direction to get the mean magnetic field $\bar{\vect{B}}$ and then subtract the mean field to get the magnetic fluctuations $\delta\vect{B}=\vect{B}-\bar{\vect{B}}$. Figure~\ref{fig:turb} shows the mean magnetic field and turbulent magnetic fluctuations. The magnitude of the mean magnetic field $\bar{B}$ shows the typical 2D flare geometry (Figure~\ref{fig:turb} (a)), including the current sheet, the Y-point, and the flare loops. $\bar{B}$ is weaker in the current sheet and near the Y-point, forming a magnetic bottle that can partially trap energetic electrons~\citep{Chen2024Energetic}. Figure~\ref{fig:turb} (b)--(d) show the three components of $\delta\vect{B}$. Overall, turbulent fluctuations can be strong locally. Due to the weak $\bar{B}$ in the CS, $\delta\vect{B}$ can be as strong as $\bar{B}$, especially above $y=120$ Mm. Most of the fluctuations in the CS are along the $y$-direction. When a flux rope is generated ($y\sim$ 130 Mm and $z>25$ Mm), $\delta\vect{B}_x$ can be strong.~\footnote{However, that could be due to the averaging procedure to get $\bar{\vect{B}}$.} It is important to note that there are only a couple of flux ropes generated in the CS due to the limited simulation resolution. We expect that the CS region could become even more turbulent in higher-resolution simulations, where more flux ropes (plasmoids in 3D) can be generated and lead to a turbulent reconnection layer~\citep{Daughton2012Emerging,Huang2016Turbulent}.

In the TS region, both $\delta B_x$ and $\delta B_y$ can be strong. The turbulence variance $\delta B^2/\bar{B}^2$ can be as large as unity locally. Such strong turbulent fluctuations can scatter particles across the shock multiple times, leading to strong particle acceleration by the TS. Figure~\ref{fig:turb} (b) shows that the largest turbulent eddies have a size of about 10 Mm (e.g., $y\sim 75-85$ Mm). Thus, it is reasonable to assume the outer scale of the turbulence inertial range or the turbulence correlation length $L_c$ is one order smaller. This finding justifies our choice of $L_c=1$ Mm for evaluating the spatial diffusion coefficients in the flare region. In the loop-top (LT) region, although the turbulence can be fairly strong, the mean field can be much stronger, as shown in Figure~\ref{fig:turb} (e). Therefore, $\delta B^2/\bar{B}^2$ is not as strong as the other regions. However, due to its broad distribution, the turbulence fluctuations can be a significant source of thermal heating and additional acceleration of the energetic particles accelerated by the CS and TS. Additionally, the SADs-induced turbulence can facilitate particle transport in the LT region, enabling particles to precipitate into lower flare loops. Recent studies of \citet{Xie2025ApJ} revealed that SADs are only one aspect of turbulent flows. It could potentially explain the broad distribution of nonthermal electrons that are responsible for the broadly distributed microwave emissions in the LT region~\citep{Chen2020Measurement}. We will come back to this point in Section~\ref{subsubsec:dist}.

\begin{figure}[ht!]
  \centering
  \includegraphics[width=\linewidth]{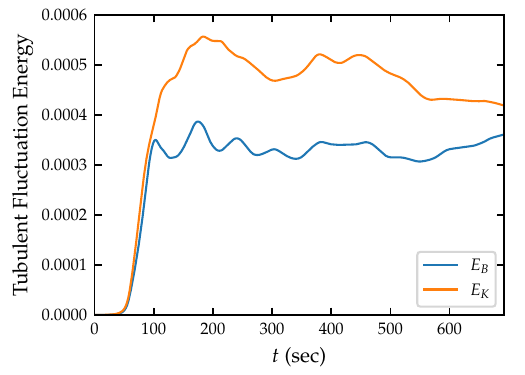}
  \caption{\label{fig:turb_ene}
  Time evolution of the turbulent fluctuation energies. $E_B=(\int \delta B^2/2) dV$ and $E_K=\int(\delta(\sqrt{\rho}V)^2/2)dV$ indicate the magnetic fluctuations and the kinetic fluctuations, respectively. The kinetic energy fluctuations are evaluated following the same procedure as the magnetic fluctuations. The values shown are in code units.
  }
\end{figure}

Figure~\ref{fig:turb_ene} shows the time evolution of the turbulent fluctuation energies. It shows that the kinetic fluctuation energy $E_K$ is even stronger than the magnetic fluctuation energy $E_B$. Both $E_B$ and $E_K$ reach their peaks early in the simulation, and they are then sustained to the end of the simulation. Thus, we expect that the strong turbulent fluctuations are continuously present in the 3D flare region, and they will facilitate particle acceleration and transport.

\subsection{Modeling electron acceleration and transport}
\label{subsec:acc}

\subsubsection{Electron spectrum}
\label{subsubsec:spect}

\begin{figure}[ht!]
  \centering
  \includegraphics[width=\linewidth]{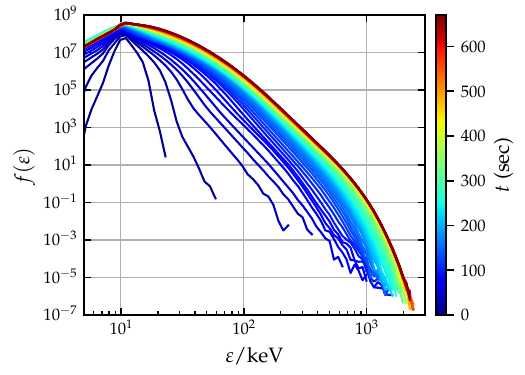}
  \caption{\label{fig:spect_global}
  Time evolution of the global electron energy spectrum for the whole domain. The time interval is $0.1T_0\approx 11$ s.
  }
\end{figure}

We then solve Parker's transport equation using the time-dependent MHD fields as background. Electrons with an initial energy of 10 keV are continuously injected into the CS. They are accelerated due to the flow compression and get transported throughout the 3D flare region. Figure~\ref{fig:spect_global} shows the time evolution of the global electron energy spectrum up to 670 s. Electrons can be accelerated to over 100 keV in the first 30 s. A large fraction of all electrons are accelerated to over 100 keV, and the maximum energy can be over 2 MeV later in the simulation. Quantitatively, about 0.2\% of all the energetic electrons in the system are over 100 keV to the end of the simulation, and the high-energy electrons ($>100$ keV) develop a power-law tail with a power-law index of $-6.5$. If we assume the plasma in the current sheet is heated to $\sim$20 MK~\citep[e.g.,][]{Warren2018ApJ,Polito2018ApJ,Chen2021Energetic,2024FrASS..1183746X}, and the injected $>$10~keV electrons represent the tail of the Maxwellian distribution, the number of electrons $>10$ keV is 0.9\% of the total electrons in the current sheet. Then, the results indicate that about $1.8\times 10^{-5}$ of all electrons in the current sheet can be accelerated to over 100 keV. Note that we have performed another simulation with a Maxwellian injection at 20 MK, and the simulation gives a similar high-energy tail, with a slightly steeper tail in the Maxwellian-injection case. Interestingly, electrons above 100 keV in the Maxwellian-injection case are $10^{-4}$ of all electrons in the current sheet, a few times higher than the mono-energetic injection case. This is because of the existing high-energy tail $> 10$ keV in the initial Maxwellian distribution (see Appendix~\ref{app:injection} for details). Such a fraction of high-energy electrons is broadly consistent with recent results derived using microwave observations of gyrosynchrotron radiation from mildly relativistic electrons. For example, in~\citet{Chen2020Measurement}, the microwave-emitting $>$300 keV electron density in the TS/LT region is found to be $10^4$--$10^5$~cm$^{-3}$, or $\sim\!10^{-5}$ of the background electron density. Joint HXR and microwave analysis that covers a broader energy regime of nonthermal electron spectra (from a few tens of keV to $\sim$MeV) returns similar results \citep{Chen2021Energetic,Chen2024Energetic}. However, the global energy spectrum cannot be used to explain these emission observations directly, which often have spatial variations. To study the acceleration and transport processes in more detail, we will look into the spatially dependent electron spectrum next and examine the electron distributions, primary acceleration mechanisms, and precipitated electrons to the flare footpoints in the following sections.

\begin{figure*}[ht!]
  \centering
  \includegraphics[width=\linewidth]{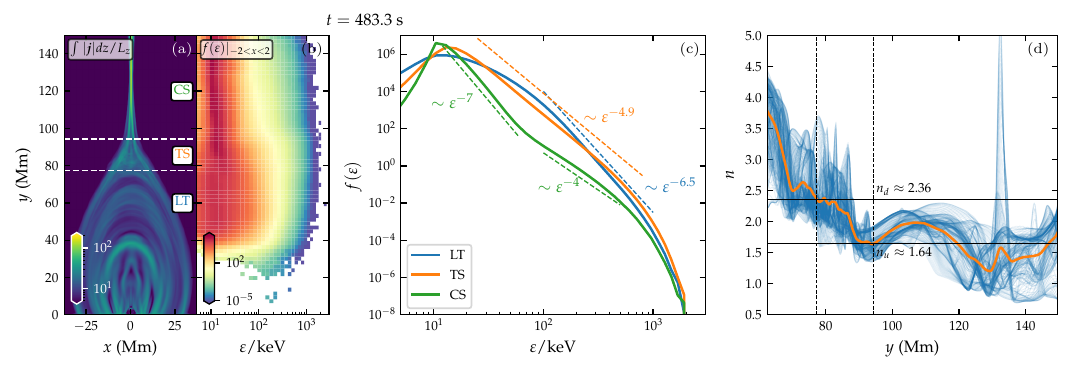}
  \caption{\label{fig:spect_cuts}
  Spatially dependent electron energy spectrum. (a) $z$--averaged current density. The white dash lines separate the reconnection, termination shock, and looptop regions. (b) The stacked electron energy spectra in the flare center (-2 Mm $< x <$ 2 Mm) along the $y$--direction. (c) The averaged electron spectra in the three regions in the flare center. They are the spectra shown (b) averaged along the $y$--direction. For the LT region, the averaging is done starting from $y=40$ Mm. The dashed lines indicate the power-law fittings. (d) The density profile in the center of the flare region ($x=0$) at $t=483.3$ s. The blue lines are 1D cuts at different $z$, and the orange curve is the average of the blue lines. The vertical dashed lines indicate the TS region, and they are at the same $y$ locations as in (a). The horizontal lines indicate the density levels in the shock upstream $n_u$ and shock downstream $n_d$. They are where the vertical dashed lines cross the orange curve.
  }
\end{figure*}

Figure~\ref{fig:spect_cuts} shows the spatially dependent electron energy spectrum. Panel (a) shows that the flare region can be roughly separated into three regions (CS, TS, and LT), as in Figure~\ref{fig:flow_comp}. Panel (b) shows the stacked electron spectra along $y$. They are obtained in the center of the flare region ($-$2 Mm $< x <$ 2 Mm). It shows the CS region has the highest low-energy fluxes because electrons are initially injected in the CS. Most of the high-energy electrons ($>100$ keV) are concentrated in the TS and LT regions. The high-energy fluxes sharply decrease towards the CS. This finding is consistent with EOVSA observations indicating that most of the high-energy electrons are in the TS/LT region instead of the CS region~\citep{Chen2020Measurement}. Some high-energy electrons ($\sim$ a few hundred keV) can diffuse into the lower flare loops ($y<40$ Mm). Figure~\ref{fig:spect_cuts} (c) shows that the spectra integrated in these three regions have distinct features. The spectrum in the CS has a double power-law feature, with low-energy steep one $\sim\varepsilon^{-7}$ and high-energy harder one $\sim\varepsilon^{-4}$. Later, we will show that the low-energy power law is primarily accelerated by the CS, while the high-energy one is due to the diffusion of energetic electrons into the CS region from the TS and LT regions. The steep low-energy power law suggests that electron acceleration in the CS is not particularly efficient. It is partly due to the lack of a large number of flux ropes, owing to the limited resolution of the MHD simulation. To investigate whether electrons can be accelerated more efficiently in the CS region, higher-resolution simulations capable of producing more flux ropes are necessary~\citep{Bhattacharjee2009Fast,Li2022Modeling}.

The energy spectrum in the TS region is an extended power law with a power-law index of $-4.9$, and it has the highest fluxes above $>300$ keV. The power-law spectrum requires a compression ratio $r\approx 1.4$ across the TS according to the diffusive shock acceleration (DSA) theory~\citep{Drury1983Introduction,Blandford1987Particle}, which gives a power-law index $3r/(r-1)$ for the momentum distribution $f(p)\sim f(\varepsilon)p^{-1}\sim p^{-10.8}$. Figure~\ref{fig:spect_cuts} (d) shows the density profile in the center of the flare region ($x=0$). It shows the compression ratio $r=n_d/n_u\approx 1.44$, where $n_u$ and $n_d$ are the densities in the TS upstream and downstream, respectively. Thus, the acceleration spectrum by the TS is consistent with what is expected from DSA. The spectrum in the LT region is similar to the global spectrum, with a lower-energy curved spectrum and a high-energy power law. Figure~\ref{fig:spect_cuts} (c) shows the high-energy flux ($>300$ keV) is in-between the ones in the CS and TS regions. However, it is important to note that Figure~\ref{fig:spect_cuts} (c) shows the averaged spectrum along the $y$-direction in each region. Since the LT is much larger, the integrated spectrum in the LT has the highest high-energy fluxes. Since the maximum energy does not change much from the TS to the LT region, the relatively steeper spectrum $>100$ keV is not caused by the high-energy acceleration. Instead, it is likely because of the low-energy acceleration in the SADs region, which boosts the low-energy fluxes and makes the high-energy tail appear steeper.

\begin{figure*}[ht!]
  \centering
  \includegraphics[width=\linewidth]{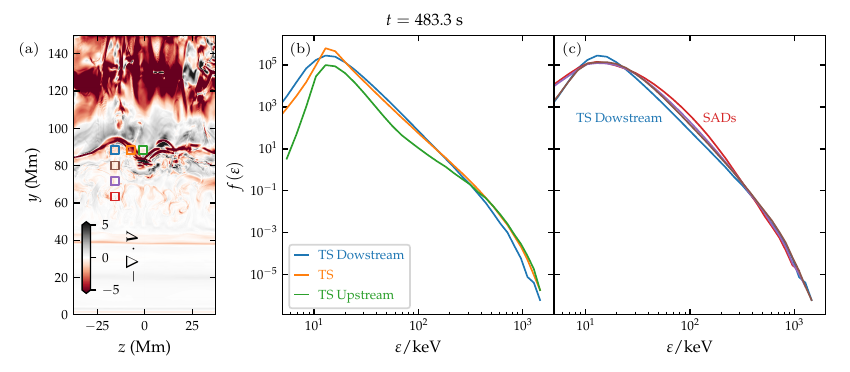}
  \caption{\label{fig:local_spect}
  Local electron energy spectra. (a) 2D slice in the $y-z$ plane (side view) of the flow compression $-\nabla\cdot\vect{V}$ overplotted with local boxes in the TS and LT regions. (b) Local electron spectra in the vicinity of the TS. The horizontally distributed boxes are at the TS downstream, TS, and TS upstream, as shown in panel (a) (c) Local electron spectra in the LT region. The boxes are vertically distributed from the TS downstream to the SADs region.
  }
\end{figure*}

To further illustrate the spatially dependent energy spectrum, Figure~\ref{fig:local_spect} shows the local electron energy spectra in the vicinity of the TS and in the LT region. The local boxes in the vicinity of the TS are in the TS downstream (blue), TS (orange), and TS upstream (green), as shown in Figure~\ref{fig:local_spect} (a). Figure~\ref{fig:local_spect} (b) shows that the spectrum at the TS and in the downstream are similar, both having an extended power-law tail. They are different at around 10--20 keV, with broader spectrum in the downstream. They are distinctly different from the spectrum in the upstream, which has a double power-law tail similar to that in the CS region (Figure~\ref{fig:spect_cuts}). Figure~\ref{fig:local_spect} (c) shows the transition from the TS downstream to the SADs region. The fluxes around 100 keV gradually increase towards the SADs, suggesting that the SADs region is efficient at accelerating electrons up to a few hundred keV. We expect the fluxes at around 100 keV to peak in the LT region due to the additional acceleration. We will further study the acceleration processes in Section~\ref{subsubsec:mechanisms}.

\subsubsection{Electron distributions}
\label{subsubsec:dist}

\begin{figure*}[ht!]
  \centering
  \includegraphics[width=0.75\linewidth]{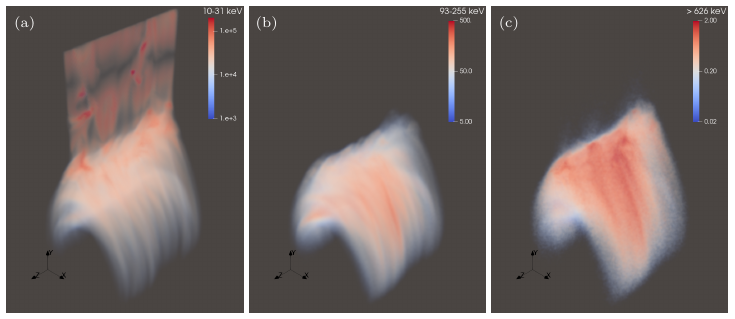}
  \caption{\label{fig:dist_3d}
  3D volume rending of the spatially dependent electron distributions at $t=417.4$ s in three energy bands (a) $10-31$ keV, (b) $93-255$ keV, and (c) $>626$ keV. There appear to be no high-energy particles in the CS region (b) and (c) primarily because of the lower bound of the colorbar range.
  }
\end{figure*}

The local spectra already show that electrons with different energies are distributed differently in the flare region. In this section, we will look into the distribution maps of the accelerated electrons. As an overview, Figure~\ref{fig:dist_3d} shows the 3D volume rendering of the spatially dependent electron distributions in three energy bands. They are not uniformly distributed along $z$ but instead form complex structures. When the distributions are mapped to the flare footpoints, we expect to see hotspots that are often observed~\citep{Liu07Eruption}. Figure~\ref{fig:dist_3d} (a) shows that low-energy (10--31 keV) electrons are broadly distributed in the entire flare region, with hotspots in the CS region and complex structures in the TS and LT regions. It is because the low-energy electrons are primarily advected with the background plasma flow, and these hotspots and structures reflect the underlying plasma structures, such as flux ropes, turbulence eddies, or SADs. Figures~\ref{fig:dist_3d} (b) \& (c) show that the high-energy (93--255 keV and $>$626 keV, respectively) electrons are mostly concentrated in the TS and LT regions. There appears to be a sharp decrease in the high-energy electron fluxes towards the CS, consistent with Figure~\ref{fig:spect_cuts} (b). Note that there are high-energy particles in the CS region, but their density is lower by several orders of magnitude and thus cannot be seen in the figure with low-density cutoffs. Their distributions in the CS will be clearer in the next two plots. Figure~\ref{fig:dist_3d} (b) shows that the particle fluxes within $93-255$ keV appear to peak below the TS and in the LT region, while the highest energy band shown in Figure~\ref{fig:dist_3d} (c) peaks right at the TS. We will examine these features later.

\begin{figure*}[ht!]
  \centering
  \includegraphics[width=\linewidth]{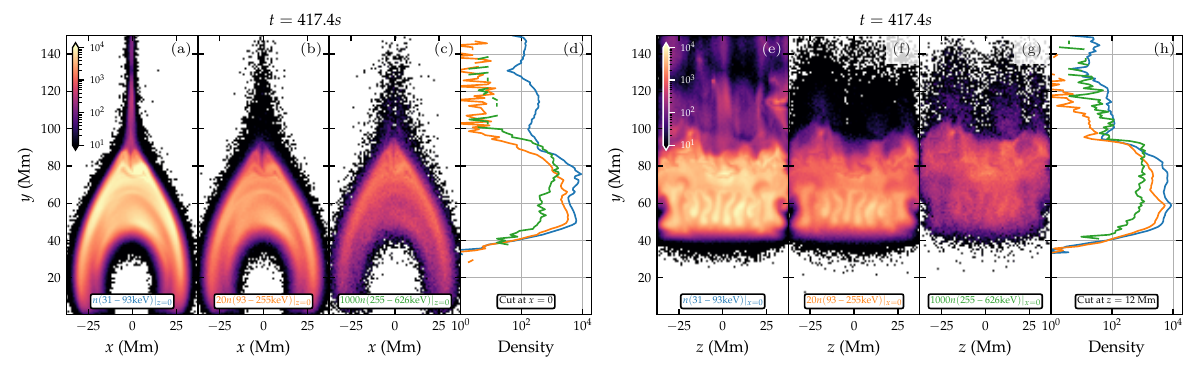}
  \caption{\label{fig:dist_2d_1d}
  2D slices and 1D cuts of the electron distributions in different energy bands. (a)--(c) 2D slices in $x-y$ plane at $z=0$. (d) 1D cuts along $x=0$ in panels (a)--(c). (e)--(g) 2D slices in $y-z$ plane at $x=0$. (h) 1D cuts along $z=12$ Mm in panels (e)--(g). Note that the two higher energy bands are multiplied by factors to make the 1D cuts at roughly the same level.
  }
\end{figure*}

To reveal the structures in electron distributions, we take 2D slices and 1D cuts of the 3D electron distributions and show the results in Figure~\ref{fig:dist_2d_1d}. Panels (a)--(c) show the 2D slices of three energy bands in the $x-y$ plane at $z=0$ (edge-on view), and panel (d) shows the 1D cuts at $x=0$ and $z=0$. Electrons in all three bands are broadly distributed in the TS/LT region and sharply decrease towards the CS region, consistent with Figure~\ref{fig:spect_cuts} (b). The 1D cuts in panel (d) show that $n(\text{31--93keV})$ in the CS region is about two orders of magnitude lower than that in the LT region (blue line), and the density jumps in the high energy bands can be even larger (orange and green lines). This finding suggests that magnetic reconnection in this system could accelerate electrons up to tens of keV but not higher-energy electrons. The 2D slices in panels (a)--(c) show that electrons are not uniformly distributed in the LT region. The 1D cuts in Figure~\ref{fig:dist_2d_1d} (d) show that $n(\text{31--93keV})$ has double peaks, one in the shock downstream and the other one at $y\approx$ 50 Mm, while $n(\text{93--255 keV})$ peaks at $y\approx$ 50 Mm. In contrast, $n(\text{255--626 keV})$ only peaks in the TS downstream, suggesting that while electrons below 255 keV could experience additional acceleration in the LT region (presumably by SADs), higher-energy electrons are not strongly affected by the plasma processes in the LT region. It becomes clearer when we look at the particle distributions from a different perspective. The 2D slices in the $y-z$ plane shown in panels (e)--(g) (side view) reveal that electron densities in the two lower energy bands are modulated by the SADs, as they have similar ``finger-like'' structures as the SADs density profiles (Figure~\ref{fig:mhd} (c)). $n(\text{255--626 keV})$ is not strongly modulated, but the 1D cut in panel (h) still shows a little peak at $y=60$ Mm (green line). The right panels of Figure~\ref{fig:dist_2d_1d} reveal another interesting feature--high-energy electrons can diffuse into the CS region from the TS region, which is particularly clear in panel (g). $n(\text{255--626 keV})$ shows enhancements around $z=-25$ Mm and $z > 0$. Comparing it with Figure~\ref{fig:mhd} (d), we find that these regions have relatively weak reconnection outflows due to 3D effects. That is why these energetic electrons can diffuse into the CS region, contributing to the high-energy tail therein (green line in Figure~\ref{fig:spect_cuts} (c)). The 1D cut at $z=12$ Mm shows that the density jump is less than two orders of magnitude (green line in panel (h)), much smaller than that in panel (d).

\begin{figure*}[ht!]
  \centering
  \includegraphics[width=\linewidth]{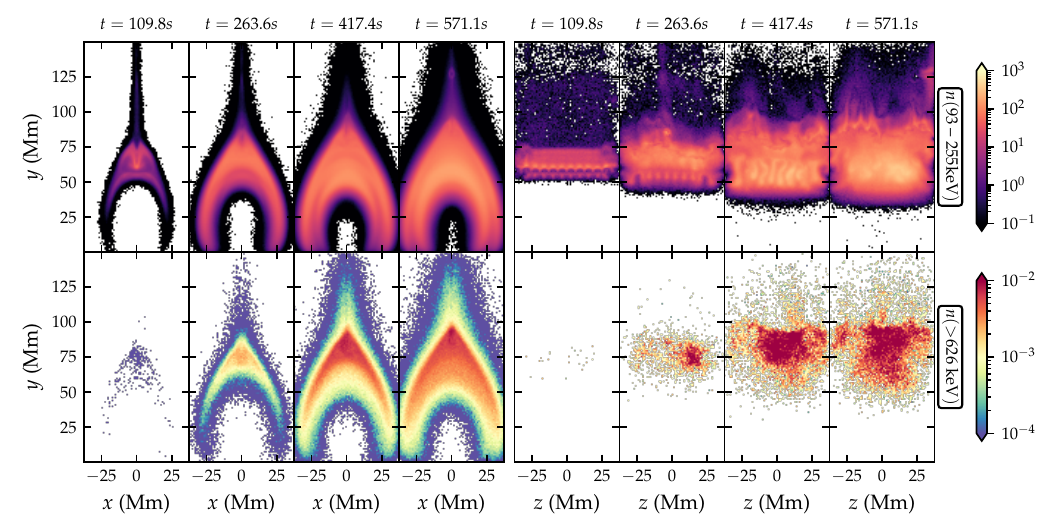}
  \caption{\label{fig:dist_time}
  Time evolution of the spatial distributions in two energy bands. The left four columns are the $z$-averaged spatial distributions. The right four columns are the 2D slices through the center of the flare region ($x=0$). The top panels are for the band 93--255 keV, and the bottom panels are for the band >626 keV.
  }
\end{figure*}

Electron distributions dynamically evolve with the MHD background. The time-evolving particle distributions can further reveal the acceleration and transport of energetic electrons in the flare region. Figure~\ref{fig:dist_time} shows the time evolution of the electron distributions in two energy bands (93--255 keV and >626 keV). These energetic electrons emerge from the TS region and gradually fill the flare loops (left panels). They gradually diffuse to lower flare loops and move upwards with the TS as reconnection proceeds (top right panels). The lower-energy band is strongly modulated by the interface instabilities where SADs appear in the LT region (top right panels), consistent with earlier results in Figure~\ref{fig:local_spect} (c) and Figure~\ref{fig:dist_2d_1d}. Later in the simulation ($t=571.1$ s), $n(\text{93--255 keV})$ peaks in the LT region due to the modulation. For the highest-energy electrons (>626 keV), their density peak in the TS downstream and above the LT region (bottom left panels). These electrons could form nonthermal emission sources in HXR and microwave bands in the above-the-looptop region~\citep{Chen2020Measurement,Chen2024Energetic}. The 2D slices in the bottom right panels show that high-energy electrons can diffuse into the lower flare loops and the CS region, consistent with earlier findings. The left panels in Figure~\ref{fig:dist_time} show that the electron density diminishes at the footpoints, especially for electrons >626 keV. It is due to the continuous particle escape at the lower boundary. These escaped electrons could produce nonthermal HXR bremsstrahlung emissions at the footpoints, and we will examine them in the next subsection.

\subsubsection{Precipitated electrons to the flare footpoints}
\label{subsubsec:precipitated}

\begin{figure*}[ht!]
  \centering
  \includegraphics[width=\linewidth]{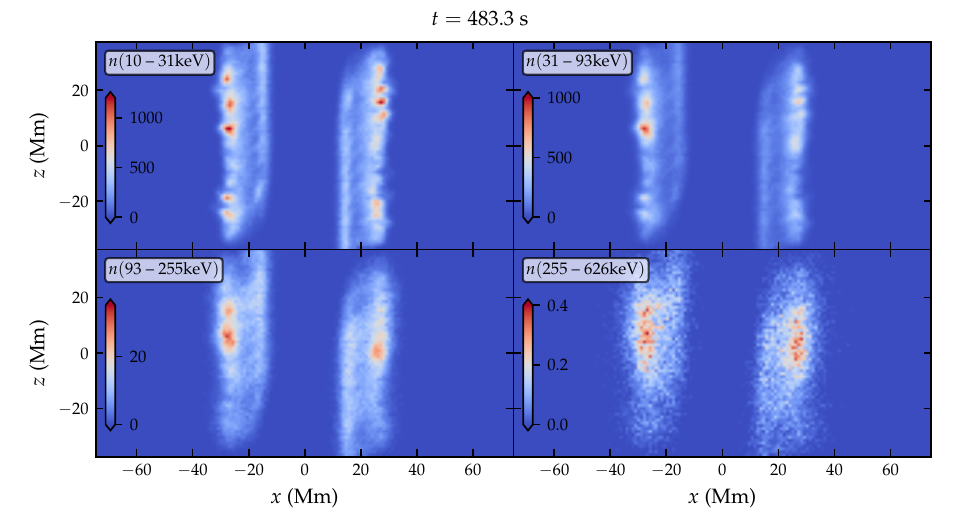}
  \caption{\label{fig:dist_footpoint}
  The distributions of the particles precipitated to the flare footpoints. These are the particles escaped from the lower boundary and accumulated for 2.2 s. Note that the colorbar range is different for different energy bands.
  }
\end{figure*}

HXR emissions serve as key diagnostics of electron acceleration in solar flares. If they are due to the bremsstrahlung process, the emission intensity is highly weighted by the local plasma density. Because plasma density is much higher at flare footpoints compared to other regions, HXR emissions are produced by the thick-target bremsstrahlung processes and usually dominate the observations. Therefore, analyzing precipitating electrons at the footpoints in our simulations offers important insights into the origins of observed HXR emissions~\citep{Liu07Eruption,Krucker11High}. Although radiation emission calculation and analysis are not included in the current work, we can approximate the electrons precipitated to the footpoints by tracking electrons that escaped from the lower boundary. Figure~\ref{fig:dist_footpoint} shows the distributions of electrons precipitated to the flare footpoints in different energy bands at $t=483.3$ s. One prompt feature is the hotspots in the distributions. These are capable of producing the hotspots in emission maps similar to observations~\citep[e.g.,][]{Liu07Eruption} if we model emissions using the spatially resolved electron distributions. The figure also reveals different features in different energy bands. There are more hotspots in the lower-energy bands (e.g., 10--31 keV). The hotspots are more compact in the lower-energy bands, and they are more diffused in the high-energy bands (e.g., 255--626 keV) due to larger spatial diffusion coefficients for high-energy electrons. Additionally, Figure~\ref{fig:dist_footpoint} shows a weak asymmetry between the two footpoints, which is likely due to asymmetric electron transport caused by the weak guide field. We expect that the asymmetry will become more pronounced in the flare region where guide field is larger, as indicated by 3D MHD simulations~\citep[e.g.,][]{Dahlin2022Variability}.

\begin{figure*}[ht!]
  \centering
  \includegraphics[width=\linewidth]{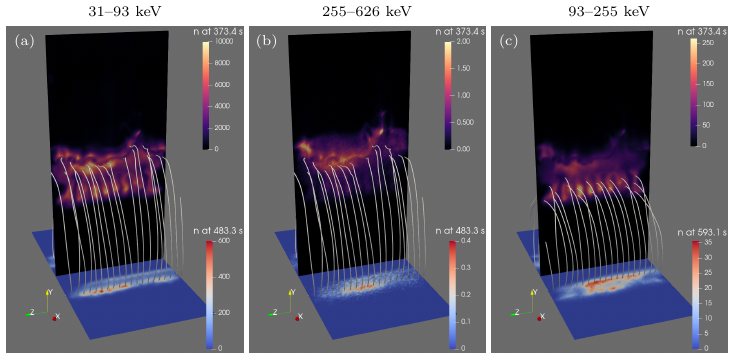}
  \caption{\label{fig:dist_mapping}
  Mapping of the electron distributions at the footpoints into the corona. The bottom planes are the distributions of the precipitated electrons, and the vertical planes indicate the 2D slices of the electron distributions at $x=0$ at earlier times. The white lines are 20 magnetic field lines traced starting from the footpoints, using magnetic fields at $t=483.3$ s for panels (a) \& (b) and $t=593.1$ s for panel (c). Their starting points are uniformly distributed along the $z$-direction. (a) Mapping of the precipitated distribution in the band of 31--93 keV at $t=483.3$ s back to the coronal electron distribution at $t=373.4$ s. (b) Similar to (a) but for the band of 255--626 keV. (c) Mapping of the precipitated distribution in the band of 93--255 keV at $t=593.1$ s back to the coronal electron distribution at $t=373.4$ s.
  }
\end{figure*}

Since the electrons displayed in Figure~\ref{fig:dist_footpoint} are transported from the coronal region, we examine the origin of these precipitated electrons by mapping the electron distributions at the footpoints into the corona, as shown in Figure~\ref{fig:dist_mapping}. Note that the vertical 2D slices indicate the coronal distributions at earlier times ($\Delta t= $ 110s, 110s, and 220s for panels (a)--(c), respectively) than that of the precipitated distributions (blue-red slices). As indicated by magnetic field connectivity, we find that some features at the footpoints can be mapped back to the TS downstream, while others can be mapped to SADs in the LT region. For example, Figure~\ref{fig:dist_mapping} (a) indicates that the hotspots in 31--93 keV are originated from the positive-$z$ region in the TS downstream. In comparison, the relatively low density in the negative-$z$ region in the precipitated distributions reflects the relatively low density in the negative-$z$ side of the TS downstream. Similarly, Figure~\ref{fig:dist_mapping} (b) shows that only part of the bright features in the TS downstream are mapped to the footpoints because some regions in the TS downstream are not magnetically connected with the footpoints. The results presented earlier in Figures~\ref{fig:local_spect} (c) and~\ref{fig:dist_2d_1d} have shown that the electron fluxes around 100 keV are enhanced by SADs. Figure~\ref{fig:dist_mapping} (c) shows that such features can be mapped to the footpoints. Since SADs are in the lower flare loops, the corresponding features at the footpoints move inwards (e.g., see cusp flare loops distribution in Figure~\ref{fig:turb} (a)) compared to the other bands. Whether the features of relative inward and diffused hotspot distribution in energy around 100-300 keV can be observed could provide critical diagnostics for electron acceleration by SADs.

\subsubsection{Primary electron acceleration mechanisms}
\label{subsubsec:mechanisms}

Previous sections have shown that electrons can be accelerated in all three acceleration regions, including CS, TS, and LT (including SADs). The results indicate that TS is the most efficient in accelerating electrons, followed by CS and LT. Could any of these mechanisms be responsible for all the high-energy acceleration? To find out the primary acceleration mechanisms, we performed additional controlled simulations, where we selectively turn on/off particle acceleration in certain regions, as indicated in Figures~\ref{fig:flow_comp} and~\ref{fig:spect_cuts}. To separate the flare regions into the three acceleration regions, we first identify the TS region based on the $V_y$ profile. First, we average $V_y$ along the $z$-direction to get $\bar{V}_y$. Second, we identify the $Y$-point, where $\bar{V}_y(x=0)$ reaches maximum along $-y$. Third, searching downwards ($-y$ direction) starting from the $Y$-point, we identify a 2D surface $\mathcal{Y}(x, z)$ where $V_y$ first reaches $-0.1V_A$, where $V_A$ is the Alfv\'en speed based on the ambient coronal plasma parameters. Such a 2D surface does not exist in the reconnection inflow plasmas, which is why, in the inflow region, it is set to be the upper boundary. Fourth, we identify the $y$-location of the TS at certain $z_0$ as the lowest point of $\mathcal{Y}(x, z_0)$, giving a TS surface uniform along $x$. Lastly, we shift the TS surface by 12 Mm upwards to the upper surface and 5 Mm downwards to get the lower surface that separates the TS region. Then, the regions above the upper surface become the CS region, and the regions below the lower surface become the LT region.

\begin{figure*}[ht!]
  \centering
  \includegraphics[width=\linewidth]{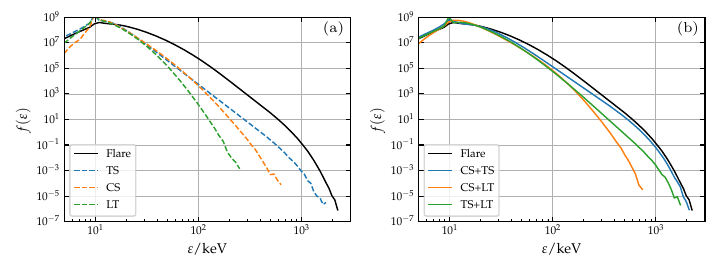}
  \caption{\label{fig:gspect_acc}
  Compare the global spectra in different control simulations. (a) The spectra in the flare simulation and the simulations with only one acceleration region. (b) The spectra in the flare simulation and the simulations with acceleration turned on in two acceleration regions.
  }
\end{figure*}

With the identified 2D surfaces, we performed 6 additional simulations with the acceleration due to flow compression being selectively turned on/off in certain regions. Three of the simulations have only one acceleration region, and the other three have two of the three acceleration regions. Figure~\ref{fig:gspect_acc} compares the global electron spectra in different simulations. In general, all six controlled simulations cannot fully reproduce the flare simulation that includes all the acceleration regions, especially the ones with only one acceleration region. Panel (a) shows that although TS itself can accelerate electrons to MeV and the high-energy tail is similar to the flare simulation, the high-energy fluxes are two orders lower than the flare simulation. Both CS and LT can accelerate electrons to a few hundred keV. Between these two, CS is slightly more efficient. However, the spectra in both simulations are significantly steeper than that in the flare simulation, and high-energy fluxes are lower than the TS simulation. Figure~\ref{fig:gspect_acc} (b) shows that a combination of two mechanisms can accelerate more electrons. Nonetheless, high-energy fluxes are still off. CS+LT (orange curve) can accelerate electrons close to MeV, but the high-energy fluxes are much lower than the flare simulation. TS+LT (green) and CS+TS (blue) simulations can accelerate electrons to over MeV because of the TS acceleration. However, although TS+LT simulation can produce similar high-energy spectral shapes, the high-energy fluxes are over one order smaller. CS+TS simulation gives a spectrum closest to the flare simulation in terms of spectral shape and maximum electron energy. The primary difference is around 100 keV, where earlier results have suggested that SADs in the LT region can modulate the acceleration and boost the fluxes around 100 keV. These results indicate that all the acceleration mechanisms play a role in accelerating a large number of electrons to high energies in the flare region. Reconnection in the CS can pre-accelerate electrons up to a few hundred keV. These electrons are then further accelerated by the TS up to a few MeV. SADs in the LT region can modulate the electron acceleration around 100 keV by boosting the fluxes. These could all produce observable signatures. In summary, it requires a synergy of different mechanisms in accelerating electrons to high energy in the flare region.

\begin{figure*}[ht!]
  \centering
  \includegraphics[width=\linewidth]{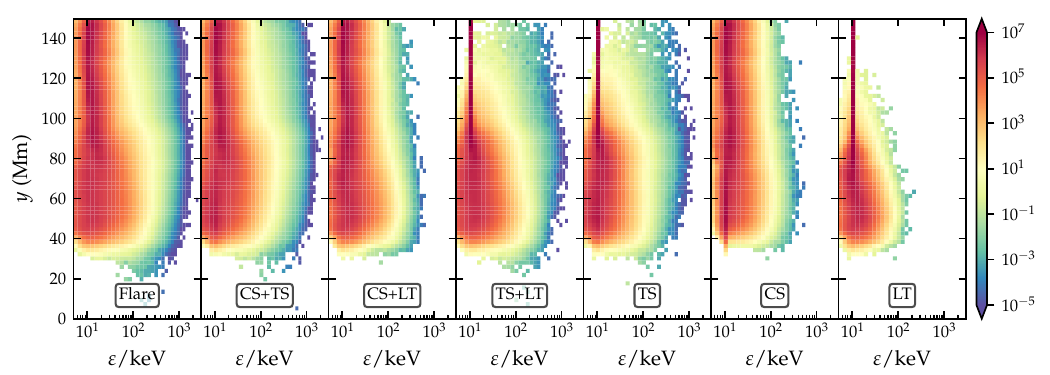}
  \caption{\label{fig:comp_stacked}
  Compare the stacked spectra in different control simulations at $t=483.3$ s. The spectra are obtained in the flare center (-2 Mm $< x <$ 2 Mm) along the $y$--direction, as in Figure~\ref{fig:spect_cuts}(b).
  }
\end{figure*}

Similar to Figure~\ref{fig:spect_cuts} (b), we can get stacked spectra in the center of the flare region along $y$ in different control simulations. Figure~\ref{fig:comp_stacked} shows the results at $t=483.3$ s. The stacked spectra in different control simulations are dramatically different from each other, and their differences further reveal the acceleration and transport processes. The CS+TS simulation is closest to the flare simulation, confirming that the LT region does not play a significant role in accelerating high-energy electrons. Similarly, the CS+LT simulation accelerates electrons to slightly higher energies than that in the CS-only simulation due to the additional acceleration in the LT region. Comparing the results in the TS+LT simulation and TS-only simulation, we find similar electron spectra in the CS region, although electron acceleration is turned off in the CS region for both cases. This result shows that high-energy electrons can diffuse into the CS region, as revealed in Figure~\ref{fig:dist_2d_1d}. Moreover, even in the LT-only case, accelerated electrons can still diffuse into the CS region along the regions with relatively weak downflows, highlighting the importance of 3D effects. In the cases with TS, electrons can be accelerated to MeV. The CS-only simulation gives relatively uniform distribution along $y$, which is consistent with earlier 2D studies~\citep{Li2022Modeling}. These 2D studies obtained similar electron distributions if magnetic islands are absent due to limited Lundquist numbers. To explain the concentrated emissions above the flare looptop region, additional acceleration mechanisms are required to boost the high-energy fluxes in the 2D simulation~\citep{Li2022Modeling}. The 2D limitation is naturally addressed in the current 3D setup, which includes multiple acceleration mechanisms. Figure~\ref{fig:comp_stacked} shows that among all the simulations, the LT-only one is the least efficient in accelerating high-energy electrons, which is expected from Figure~\ref{fig:flow_comp} (c) showing the weakest flow compression in the SADs region.

\begin{figure*}[ht!]
  \centering
  \includegraphics[width=\linewidth]{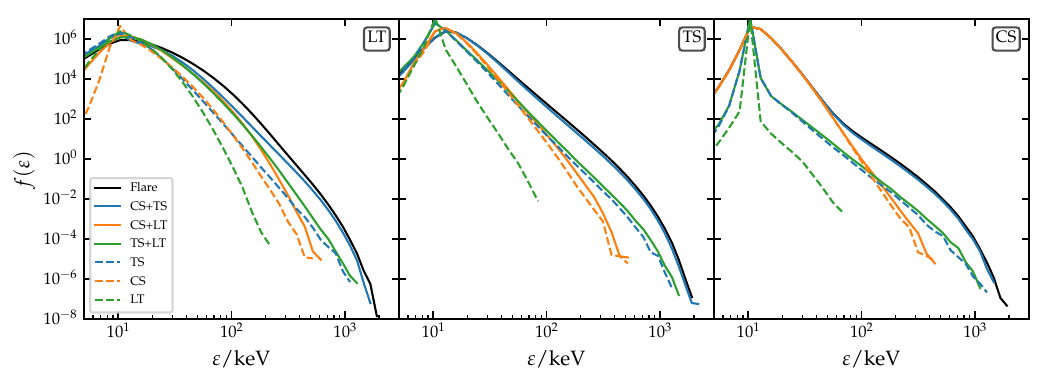}
  \caption{\label{fig:comp_local}
  Compare the local spectra in different control simulations at $t=483.3$ s. The spectra are obtained in the flare center (-2 Mm $< x <$ 2 Mm). They are accumulated in the three acceleration regions, as in Figure~\ref{fig:spect_cuts}(c). The colors and line types are the same as in Figure~\ref{fig:gspect_acc}.
  }
\end{figure*}

Similar to Figure~\ref{fig:spect_cuts} (c), we can accumulate the electron spectra in the three acceleration regions. Figure~\ref{fig:comp_local} shows the results at $t=483.3$ s. Since most of the energetic electrons are concentrated in the LT region, the conclusions found in Figure~\ref{fig:gspect_acc} apply to the LT region (left panel). In the TS region (middle panel), the CS+TS simulation (blue solid line) gives an almost identical spectrum as the flare simulation, suggesting that the LT region plays a minimum role in the TS acceleration. This is reasonable because particles are initially injected into the CS region and transport downwards. The electron fluxes in TS+LT case (green solid line) are much lower than that in the flare simulation or the CS+TS simulation, demonstrating the importance of the pre-acceleration in the CS region. The TS+LT and TS-only (blue dashed line) simulations give similar electron spectra in the TS region, and the CS+LT (orange solid line) and CS-only (orange dashed line) simulations give similar spectra, both of which suggest a minimum role of the LT-acceleration in the TS region. The green dashed line indicates that electrons accelerated in the LT region can diffuse into the TS region, though the fluxes are very low. In the CS region (right panel), the same conclusions were found in the TS region. Comparing the orange curves with the black and blue curves, we find that the high-energy tails are originated from the TS and LT regions. It does not mean that the CS region can be excluded. The electron fluxes are much lower when electron acceleration is turned off in the CS region (green solid line and blue dashed line), suggesting again that the pre-acceleration in the CS region is critical to boosting the electron fluxes. The green dashed line shows that very few particles can reach the CS region in the LT-only simulation. These results show that each of the mechanisms cannot reproduce the spectra and spatial distributions of the flare simulation that includes multiple acceleration mechanisms. The synergy of different mechanisms in accelerating electrons highlights the importance of particle transport between different regions.

\begin{figure*}[ht!]
  \centering
  \includegraphics[width=\linewidth]{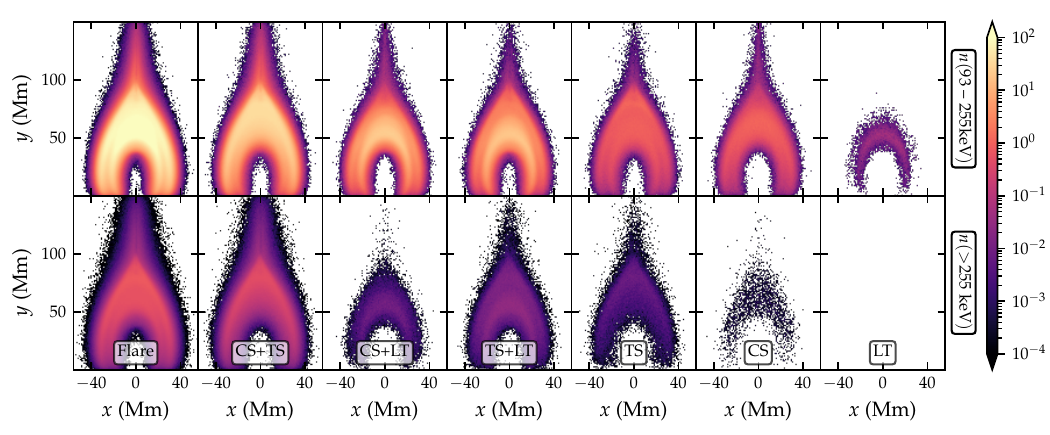}
  \caption{\label{fig:comp_dist}
  Compare electron distribution maps in different control simulations in the $x-y$ plane at $t=483.3$ s. The top panels are in the energy band of $93-255$ keV, and the bottom panels are in the energy band of $>255$ keV.
  }
\end{figure*}

The 2D distribution maps can further reveal the dramatic differences between different simulations. Figure~\ref{fig:comp_dist} shows the results in two energy bands at $t=483.3$ s. In the lower-energy band (top panels), although the morphology of the distribution maps is similar (except for the LT-only simulation), all the controlled simulations underestimate the electron fluxes. The CS+TS simulation is the closest to the flare simulation. In the high-energy band, the CS+TS simulation produces similar morphology and electron fluxes, and the rest strongly underestimate the high-energy electron fluxes. Therefore, we expect that the CS+TS simulation could produce similar emission maps in the microwave bands observed by EOVSA~\citep{Chen2020Measurement}.

\begin{figure*}[ht!]
  \centering
  \includegraphics[width=\linewidth]{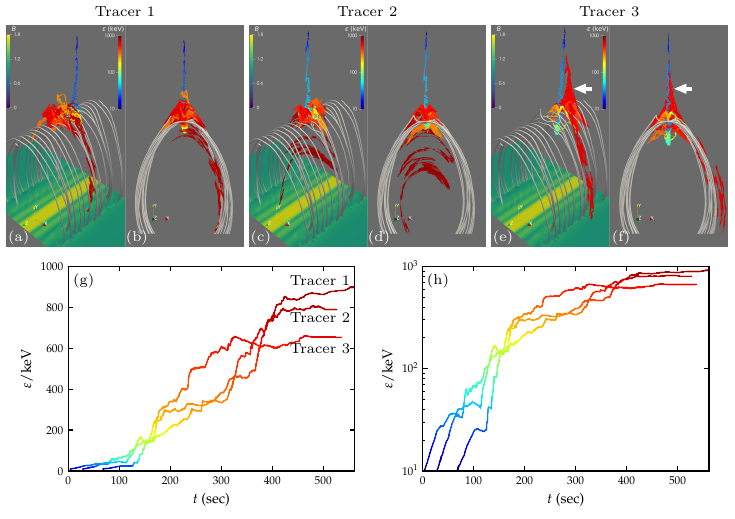}
  \caption{\label{fig:ptl_traj}
  Tracer trajectories of three pseudo particles. The top panels show the 3D views and 2D projections of the tracer trajectories, which are color-coded by the particle kinetic energies. The white lines are sample magnetic field lines traced using the magnetic fields at $t=483.3$ s. The magnetic field strength $B$ at the lower boundary is also shown. All three particles precipitated to the footpoints at the end of the simulation. (g) Time evolution of the particle kinetic energy $\varepsilon$ in linear scale. (h) Time evolution of $\varepsilon$ in logarithmic scale. The lines are color-coded by particle kinetic energies similar to the top panels.
  }
\end{figure*}

Although we use pseudo particles instead of physical particles to solve the transport equation, their trajectories are still useful for revealing the acceleration and transport processes. Figure~\ref{fig:ptl_traj} shows three tracer particle trajectories. All three pseudo-particles are accelerated to a few hundred keV. Tracer 1 represents the behavior of most high-energy particles. It is initially accelerated in the CS region and gets advected downwards by the reconnection outflow. It gains most of its kinetic energy near the TS region, and its trajectory is quite complex in the TS region due to the turbulent background. After being accelerated, it then diffuses to the flare footpoint. Tracer 2 is similar to Tracer 1 in many ways. The difference is that turbulence induced by SADs in the LT region helps the particle transport to the lower flare loops, enabling the particle to reach the inner region of the flare footpoints. Tracer 3 is more unique because it is accelerated first but then diffuses into the CS during the late stage of the simulation (after $t\approx 350$; indicated by the white arrows in panels (e) and (f)), contributing to the high-energy tail of the electron spectra in the CS region (e.g., Figure~\ref{fig:spect_cuts} (c)). Figure~\ref{fig:ptl_traj} (g) \& (h) shows the time evolution of the particle kinetic energies in linear scale and logarithmic scale, respectively. The linear plot shows that the energy gain in the CS region is minimal compared to the energy gain afterward. However, the logarithmic plot reveals that the energy gains in the CS region are as fast as the following acceleration. It is due to the nature of Fermi acceleration, which gives $\dot{p}\sim p$ or $\dot{\varepsilon}\sim\varepsilon$. The pre-acceleration in the CS region becomes critical for more efficient following acceleration in the TS region.

\section{Discussions and Conclusions}
\label{sec:con}

In this paper, we have studied electron acceleration and transport in 3D solar flare regions by solving Parker's transport equation in a 3D MHD simulation with a standard eruptive flare geometry that features a vertical current sheet above reconnected flare arcades ~\citep{Shen2022}. The 3D MHD simulation has been demonstrated to include several types of particle acceleration mechanisms~\citep{Shen2022}. We have examined the energy conversion, flow compression distribution, and plasma turbulence in the 3D system. We then solved Parker's transport equation using the 3D fields from the MHD simulation to study electron acceleration and transport. Specifically, we have examined the electron energy spectra, electron spatial distribution maps, precipitated electrons to the flare footpoints, and primary electron acceleration mechanisms. We find the following conclusions.

\begin{itemize}
  \item By examining the energy conversion in the system, we find that although magnetic energy release is driven by magnetic reconnection in the CS region, the released energy is primarily converted into bulk kinetic energy. The increase of the internal energy occurs throughout the flare region due to flow compression, but it is most efficient in the TS region. The distribution of the flow compression also reveals that it is the strongest near the TS, followed by the CS and SADs. Based on these findings, we expect that the particle acceleration due to flow compression is most efficient at the TS, followed by the CS and LT (including SADs) regions.
  \item Large-amplitude turbulence fluctuations are generated and sustained throughout the 3D flare region. The turbulence variance $\delta B^2/\bar{B}^2$ can be as large as 1 locally, especially in the CS and TS regions. The largest turbulence eddies have a size of about 10 Mm, suggesting the turbulence correlation length $L_c\sim$ 1 Mm. These findings justify our choice of turbulence parameters when solving the transport equation.
  \item By solving Parker's transport equation with the 3D MHD fields as background, we find that a large number of electrons are accelerated to high energies and develop a power-law tail with a spectral index of $-$6.5 in the global energy spectrum. Specifically, we estimate that $1.8\times 10^{-5}$ of all electrons in the current sheet/looptop region can be accelerated to over 100 keV, which is consistent with recent observational results based on microwave and HXR spectral diagnostics~\citep{Chen2020Measurement,Chen2021Energetic,Chen2024Energetic}.
  \item Local electron spectra show distinct features in different regions. The energy spectrum in the CS region has a double power-law feature with spectral indices of $-$7 and $-$4. The spectrum is a single power law in the TS region, and the spectral index of $-$4.9 is consistent with diffusive shock acceleration. The spectrum in the LT region is similar to the global spectrum, which has a high-energy power-law tail and a bump of around 100 keV.
  \item Both the local spectra and spatial distributions reveal that most of the high-energy electrons are concentrated in the TS and LT regions, and they sharply decrease towards the CS region. Low-energy electrons ($<$100 keV) are broadly distributed throughout the flare region, as they are primarily advected by the background plasma flows. The fluxes around 100 keV peak in the LT region due to the modulation of the SADs of these electrons, as there is weak flow compression in the SADs region. The fluxes for the highest energies (> a few hundred keV) peak in the TS region downstream due to the acceleration by the TS. Additionally, we find that energetic electrons ($>100$ keV) can diffuse into the CS region from the TS and LT regions, contributing to the high-energy tail in the CS region.
  \item By tracking particles escaping from the lower boundary, we can approximate the precipitated electrons to the flare footpoints. We find that the precipitated electrons develop hotspots along the footpoints, and the features are different in different energy bands. These hotspots at the footpoints can be mapped back into the coronal sources in the TS and SADs regions at earlier times.
  \item By selectively turning on/off the acceleration in certain regions (CS, TS, and LT), we can pinpoint the acceleration and transport processes. We find a synergy of these mechanisms in accelerating a large number of electrons to high energies. TS is responsible for the acceleration of MeV electrons, and CS and LT can accelerate electrons up to a few hundred keV. Each of the individual mechanisms cannot reproduce the spectra and spatial distributions of the flare simulation that includes all mechanisms. In the 3D system, electrons are pre-accelerated in the CS region, and they are further accelerated in the TS region. These accelerated electrons are transported in the LT region and get modulated in the SADs region, especially around 100 keV.
\end{itemize}

The model is capable of producing spatially and temporally dependent electron energy spectra and distribution maps in different energy bands in 3D. Although these can reveal the acceleration and transport processes, to directly compare with remote-sensing observations from different viewing perspectives, it is necessary to use the spectra and distribution maps to model the thermal and nonthermal emissions at multiple wavelengths. Our preliminary results have shown that the distribution maps of the precipitated electrons can produce localized HXR footpoint sources, a phenomenon often observed in two-ribbon flares \citep[e.g.,][]{LiuC2007}. Despite its strengths, the current model lacks key physics such as thermalization and collisional energy losses~\citep{Fletcher1995AA,Holman2011,Battaglia2012,Jeffrey2014PhDT}. These effects are especially significant in the high-density looptop region and flare loops, where particle collisions can be frequent enough to drive collisional energy loss and affect particle transport in the regions. For example, the frequent collisions could lead to more trapping in the looptop region and slower precipitation to the flare footopints~\citep{Kong2022Numerical}. Expanding upon our recent work based on a 2.5D model that combines MHD simulation, particle modeling, and synthetic emission generation \citep{Chen2024Energetic}, we are actively working on a 3D model of the radiation emission, including the radiation transfer, for direct comparison with HXR and microwave emission observations from different viewing perspectives. We will defer these studies for future publications.

Our results address some issues of earlier 2D studies if plasmoids are absent~\citep[e.g.,][]{Li2022Modeling}, which often show that the high-energy fluxes do not peak in the LT region as observed~\citep{Chen2020Measurement}. Because of the additional acceleration by TS and SADs and particle trapping by SADs-induced turbulence, high-energy fluxes strongly peak in the TS and LT regions. Such locally enhanced electron fluxes are capable of explaining the HXR and microwave emissions in the LT and above-the-LT region.

The modeling results indicate that the acceleration is weak in the CS region, which produces a spectrum with a power-law index of $-7$. It is likely due to the limited resolution of the MHD simulation with a relatively low Lundquist number of $5\times 10^4$ globally, which prevents the generation of a large number of flux ropes~\citep[or plasmoids in 2D,][]{Samtaney2009,Huang2010Scaling,Loureiro2012}. These flux ropes will help convert the bulk flow energy into plasma internal energy and form more compression layers in the CS region, leading to more efficient electron acceleration. However, it is important to note that the modeling result is actually consistent with flare observations, where large magnetic islands and localized nonthermal emissions are rarely observed within the flare current sheet. Moreover, our MHD simulation shows no plasmoid mergers, consistent with those 2.5D flare simulations that used realistic geometries and boundary conditions~\citep[e.g.,][]{Karpen2012ApJ,Guidoni2016ApJ}. The question is whether higher-resolution simulations will give more efficient electron acceleration in the CS region and, at the same time ``hide'' the flux ropes from observations. It is possible since the flux ropes in a 3D reconnection layer are subject to plasma instabilities (e.g., kink and coalescence), and most of them will be disrupted before they can grow to large sizes. The resulting reconnection layer becomes turbulent~\citep{Daughton2011Role,Dahlin2015Electron,Li2019Formation,Guo20213D,Zhang2021Efficient}, and that could explain the absence of observable flux ropes/plasmoids in imaging observations~\citep{Cheng2018ApJ,Warren2018ApJ,French2019}.

Besides the Lundquist number, other MHD simulation parameters, such as plasma $\beta$ and guide field, can also affect particle acceleration and transport. These should be modeled more accurately in MHD simulations to study electron acceleration in solar flares. These plasma parameters strongly affect flow compression and hence particle acceleration~\citep{Li2018Roles}. Typically, plasmas can be compressed more easily in the lower-$\beta$ and lower guide-field reconnection layers, where more efficient acceleration is expected. More interestingly, both MHD simulations~\citep{Dahlin2022Variability} and flare observations~\citep{Qiu2023Role} have indicated that the guide field is not a constant but keeps evolving during a solar flare event. How the time-varying guide field affects electron acceleration and transport requires future studies. Additionally, to compare with in-situ observations by PSP and Solar Orbiter and current/future radio/HXR observations with higher dynamic range and/or field of view~(e.g., OVRO-LWA; \citealt{Anderson2018}, FASR; \citealt{FASR_WP}, and a spacecraft version of FOXSI; \citealt{Buitrago-Casas2021}), it is necessary to perform MHD simulations and particle modelings to include processes like flux rope ejection and CME formation.

In this study, we use Parker's transport equation that assumes a nearly isotropic particle distribution. It could be reasonable if reconnection-driven turbulence~\citep{Daughton2011Role,Guo2015Efficient,Dahlin2015Electron,Li2019Formation,Zhang2021Efficient} or instabilities~\citep{Roberg-Clark2019ApJ} can strongly scatter the energetic particles. In this regime, flow compression is the first-order acceleration mechanism in the reconnection layer, as demonstrated by earlier kinetic simulations~\citep{Li2018Roles}. Importantly, its overall effect is equivalent to that produced by guiding-center drift acceleration~\citep{Dahlin2014Mechanisms,LeRoux2015Kinetic,Li2017Particle,Li2018Roles}. However, if the pitch-angle scattering is insufficient to maintain isotropic particle distributions, we need to evolve particle pitch angles together with particle momentum. We can solve either the focused transport equations~\citep{Zank2014Particle,LeRoux2015Kinetic} or the drift kinetic equation~\citep{Drake2013Power,Montag2017Impact}. Additionally, it is necessary to account for acceleration mechanisms that become relevant when particle distributions are anisotropic. For example, flow shear acceleration can play a significant role under these conditions~\citep{Earl1988Cosmic,LeRoux2015Kinetic,Webb2018,Webb2019,Li2018Roles}.

One of the mechanisms that is not specifically included in our model is the parallel electric field $E_\parallel$, which could play a significant role in particle acceleration in solar flares. Kinetic simulations show that $E_\parallel$ is important for injecting particles out of the thermal background in both reconnection and turbulence, with its role becoming more important in strong guide-field cases~\citep[e.g.,][]{Dahlin2014Mechanisms,French2023ApJ,Zhang2024Plasma}. Since Parker's transport equation does not address particle injection, this effect is not explicitly modeled. Instead, we approximate it by injecting nonthermal particles in regions with strong current densities, such as current sheets. Additionally, $E_\parallel$ along exhaust boundaries can form electrostatic potential wells~\citep{Egedal2013Review}, which help trap electrons and facilitate Fermi acceleration. This effect involves strong anisotropies and cannot be captured by Parker's equation, which assumes isotropic particle distributions. More advanced models, such as the focused transport equation~\citep{LeRoux2018Self} or guiding-center-based approaches~\citep[e.g., the kglobal model by][]{Drake2019,Arnold2021PRL,Yin2024Simultaneous} are required to incorporate this physics.

While first-order effects depend strongly on the particle distribution anisotropy, as discussed earlier, it is also important to consider second-order acceleration processes arising from fluctuations in $E_\parallel$, flow compression, and shear. In a turbulent 3D flare environment, these fluctuations can induce stochastic acceleration even when pitch-angle anisotropy is small~\citep{LeRoux2018Self}. To simplify the problem and build a baseline, we have so far included only first-order acceleration due to flow compression. Incorporating second-order effects is essential for a more complete and realistic description of particle acceleration and transport, which we aim to pursue in future work. It is worth noting that the 3D flare simulation exhibits a broad distribution of flow compression (Figure~\ref{fig:flow_comp} (c)). As a result, particles experience varying acceleration depending on their location, which leads to a spread in the energy distribution. In this sense, some level of fluctuation-driven effects is implicitly captured. However, our current model does not resolve sub-grid-scale fluctuations and therefore cannot fully account for stochastic acceleration as described in the perturbation approach~\citep{LeRoux2018Self}.

Another potential caveat of the current study is that the feedback from energetic particles to the background plasma is not included. This assumption should be valid if the nonthermal fraction is small. In our case, we assume the initially injected 10 keV electrons are from the high-energy tail of the strongly heated plasma in the current sheet. For a 20 MK plasma, the fraction of 10 keV electrons is only 0.9\%. Although these particles could affect the dynamics of the background thermal plasma, such effects should be minimal according to a recent study that includes feedback effects~\citep {Seo2024Proton}. However, some studies suggest that, in certain flares, the nonthermal electron density may be comparable to that of the ambient plasma~\citep{Krucker2010Measure,Fleishman2022Natur}. In these cases, the nonthermal particles could play a significant role in modifying the reconnection dynamics and changing the particle acceleration/transport processes in flares. Accordingly, the feedback of energetic particles through energy (or pressure) exchange or current density should be included in the model~\citep{Bai2015ApJ,Drake2019,Arnold2019,Arnold2021PRL,Seo2024Proton,Yin2024Simultaneous}. Additionally, the choice of injected particle energy (e.g., 10 keV) is not fully self-consistent. A more accurate treatment of particle injection requires fully kinetic simulations~\citep[e.g.,][]{Zhang2021Efficient} to determine how particles are energized from the thermal background.

Lastly, there are additional uncertainties in the turbulence properties and particle scattering mechanisms. First, although we have a rough idea of the turbulence properties in the flare region, including turbulence amplitude and correlation length, additional studies are still required to understand the properties of the reconnection-driven turbulence~\citep{Huang2016Turbulent,Kowal2017Statistics,Daughton2011Role,Li2019Formation,Guo20213D}, SADs-driven turbulence~\citep{Xie2025ApJ}, and turbulence in the vicinity of termination shocks. According to current theories, plasma turbulence is typically anisotropic~\citep[e.g.,][]{GS95}, with much more wave power in $k_\perp$ than $k_\parallel$ wave modes, which has been confirmed by space plasma observations~\citep[e.g.,][]{Chen2016JPlPh}. Since particles primarily interact with the $k_\parallel$ wave modes, turbulence anisotropy strongly affects particle scattering and transport. We expect that the progress on turbulence properties in the flare region can lead to more accurate spatial or energy diffusion coefficients than estimates using the quasi-linear theory. Second, the turbulence properties could vary in different regions. Particle modeling with spatially dependent turbulence properties will likely lead to better comparisons with flare observations. Third, the mechanisms of electron scattering in solar flares are not well established. Given the large number of accelerated electrons, self-generated waves (e.g., whistlers) may play a significant role, but their relative importance compared to background or reconnection-driven turbulence is unclear. To avoid these complexities, we adopted an isotropic and spatially unifor turbulence model as a first step, as it provides the simplest framework and is commonly used in quasi-linear theories of particle scattering and diffusion~\citep[e.g.,][]{Jokipii1971Propagation,Giacalone1999Transport}. We plan to incorporate more realistic turbulence physics in future work to enable more direct comparisons with observations.

To conclude, by solving Parker's transport equation using 3D MHD simulation results, we model the electron acceleration and transport in the 3D flare region. We find that a significant number of electrons can be accelerated to high energies through the synergy of different acceleration mechanisms, including reconnecting the current sheet, the termination shock, and supra-arcade downflows. The model produces spatially and temporally dependent electron energy spectra and spatial distributions that are broadly consistent with recent observational results, and can be further utilized to compare with multi-wavelength observations from different perspectives.

\begin{acknowledgments}
We thank the supports from NASA through grant NNH240B72A and NSF grants No. AST-2107745 and AGS-2334930. X.L. acknowledges the support from NASA LWS grant 80NSSC21K1313, University of California, Berkeley through subcontract No. 00011753, and Smithsonian Astrophysical Observatory through subcontract No. SV1-21012. C.S. and X.X. are supported by NSF grant AGS-2334929. F.G. acknowledges the support from NASA grants 80HQTR21T0087 and 80HQTR21T0104 and NSF grant AST-2109154. B.C., I.O., Y.W., and S.Y. acknowledge additional support from NASA grant 80NSSC20K1318 and NSF grants AST-2108853 and AGS-2334931 to NJIT. This research used resources of the National Energy Research Scientific Computing Center (NERSC), a Department of Energy Office of Science User Facility using NERSC award FES-ERCAP0033004.
\end{acknowledgments}

\appendix
\section{Compare particle injection methods}
\label{app:injection}

\begin{figure}[ht!]
  \centering
  \includegraphics[width=0.5\linewidth]{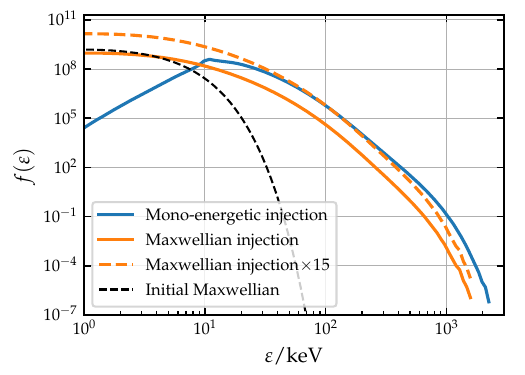}
  \caption{\label{fig:injection}
  Final electron spectra for mono-energetic injection at 10 keV and Maxwellian injection at 20 MK. Since the number of injected macro particles is the same in both cases, $\int f(\varepsilon)d\varepsilon$ is similar. The orange dashed line shows the shifted spectrum for the Maxwellian injection. The black dashed line indicates the initial Maxwellian distribution.
  }
\end{figure}

To illustrate the effect of particle injection, we performed another simulation, where the injected particles have a Maxwellian distribution with a thermal temperature of 20 MK to mimic the super-hot plasma in the flare region. There are about 0.9\% of electrons with energies $\geq 10$ keV in the Maxwellian distribution. The number of injected macro particles is the same in the both cases. Figure~\ref{fig:injection} compares the final electron spectra in the two cases. The results show that the spectral shapes at high energies are similar between the two cases. However, because the same number of macro particles is injected in both, the Maxwellian-injection case yields a significantly lower $f(\varepsilon > 10\mathrm{keV})$ compared to the mono-energetic injection. At $\varepsilon = 1000$ keV, the difference reaches two orders of magnitude, consistent with that only about 0.9\% of electrons in a Maxwellian distribution exceed 10 keV. In the Maxwellian-injection case, most electron flux is at lower energies ($<10$ keV), which dominates computational cost without contributing to nonthermal acceleration. The resulting low fluxes of high-energy particles also makes it challenging to obtain statistically robust electron distribution maps at high energies. For these reasons, mono-energetic injection is adopted in this study. Interestingly, there are about $10^{-4}$ electrons above 100 keV in the Maxwellian-injection case, which is a few times higher than $1.8\times 10^{-5}$ in the mono-energetic-injection case. The reason is that the injected electrons in the Maxwellian-injection case have significant fluxes above 10 keV (black dashed line).


\begin{thebibliography}{}
\expandafter\ifx\csname natexlab\endcsname\relax\def\natexlab#1{#1}\fi
\providecommand{\url}[1]{\href{#1}{#1}}
\providecommand{\dodoi}[1]{doi:~\href{http://doi.org/#1}{\nolinkurl{#1}}}
\providecommand{\doeprint}[1]{\href{http://ascl.net/#1}{\nolinkurl{http://ascl.net/#1}}}
\providecommand{\doarXiv}[1]{\href{https://arxiv.org/abs/#1}{\nolinkurl{https://arxiv.org/abs/#1}}}

\bibitem[{{Adhikari} {et~al.}(2019){Adhikari}, {Khabarova}, {Zank}, \&
  {Zhao}}]{Adhikari2019Role}
{Adhikari}, L., {Khabarova}, O., {Zank}, G.~P., \& {Zhao}, L.~L. 2019, \apj,
  873, 72, \dodoi{10.3847/1538-4357/ab05c6}

\bibitem[{{Anderson} {et~al.}(2018){Anderson}, {Hallinan}, {Eastwood},
  {Monroe}, {Vedantham}, {Bourke}, {Greenhill}, {Kocz}, {Lazio}, {Price},
  {Schinzel}, {Wang}, \& {Woody}}]{Anderson2018}
{Anderson}, M.~M., {Hallinan}, G., {Eastwood}, M.~W., {et~al.} 2018, \apj, 864,
  22, \dodoi{10.3847/1538-4357/aad2d7}

\bibitem[{{Arnold} {et~al.}(2019){Arnold}, {Drake}, {Swisdak}, \&
  {Dahlin}}]{Arnold2019}
{Arnold}, H., {Drake}, J.~F., {Swisdak}, M., \& {Dahlin}, J. 2019, Physics of
  Plasmas, 26, 102903, \dodoi{10.1063/1.5120373}

\bibitem[{{Arnold} {et~al.}(2021){Arnold}, {Drake}, {Swisdak}, {Guo}, {Dahlin},
  {Chen}, {Fleishman}, {Glesener}, {Kontar}, {Phan}, \& {Shen}}]{Arnold2021PRL}
{Arnold}, H., {Drake}, J.~F., {Swisdak}, M., {et~al.} 2021, \prl, 126, 135101,
  \dodoi{10.1103/PhysRevLett.126.135101}

\bibitem[{{Aschwanden} {et~al.}(2016){Aschwanden}, {Holman}, {O'Flannagain},
  {Caspi}, {McTiernan}, \& {Kontar}}]{Aschwanden2016ApJ}
{Aschwanden}, M.~J., {Holman}, G., {O'Flannagain}, A., {et~al.} 2016, \apj,
  832, 27, \dodoi{10.3847/0004-637X/832/1/27}

\bibitem[{{Bai} {et~al.}(2015){Bai}, {Caprioli}, {Sironi}, \&
  {Spitkovsky}}]{Bai2015ApJ}
{Bai}, X.-N., {Caprioli}, D., {Sironi}, L., \& {Spitkovsky}, A. 2015, \apj,
  809, 55, \dodoi{10.1088/0004-637X/809/1/55}

\bibitem[{{Battaglia} {et~al.}(2012){Battaglia}, {Kontar}, {Fletcher}, \&
  {MacKinnon}}]{Battaglia2012}
{Battaglia}, M., {Kontar}, E.~P., {Fletcher}, L., \& {MacKinnon}, A.~L. 2012,
  \apj, 752, 4, \dodoi{10.1088/0004-637X/752/1/4}

\bibitem[{{Bhattacharjee} {et~al.}(2009){Bhattacharjee}, {Huang}, {Yang}, \&
  {Rogers}}]{Bhattacharjee2009Fast}
{Bhattacharjee}, A., {Huang}, Y.-M., {Yang}, H., \& {Rogers}, B. 2009, Physics
  of Plasmas, 16, 112102, \dodoi{10.1063/1.3264103}

\bibitem[{{Birn} {et~al.}(2009){Birn}, {Fletcher}, {Hesse}, \&
  {Neukirch}}]{Birn2009Energy}
{Birn}, J., {Fletcher}, L., {Hesse}, M., \& {Neukirch}, T. 2009, \apj, 695,
  1151, \dodoi{10.1088/0004-637X/695/2/1151}

\bibitem[{{Blandford} \& {Eichler}(1987)}]{Blandford1987Particle}
{Blandford}, R., \& {Eichler}, D. 1987, \physrep, 154, 1,
  \dodoi{10.1016/0370-1573(87)90134-7}

\bibitem[{{Buitrago-Casas} {et~al.}(2021){Buitrago-Casas}, {Vievering},
  {Musset}, {Glesener}, {Athiray}, {Baumgartner}, {Bongiorno}, {Champey},
  {Christe}, {Courtade}, {Dalton}, {Duncan}, {Gilchrist}, {Ishikawa},
  {Jhabvala}, {Kanniainen}, {Krucker}, {Gregory}, {Martinez Oliveros},
  {McCracken}, {Mitsuishi}, {Narukage}, {Pantazides}, {Peretz}, {Perez-Piel},
  {Ramanayaka}, {Ramsey}, {Ryan}, {Savage}, {Takahashi}, {Watanabe},
  {Winebarger}, \& {Zhang}}]{Buitrago-Casas2021}
{Buitrago-Casas}, J.~C., {Vievering}, J., {Musset}, S., {et~al.} 2021, in
  Society of Photo-Optical Instrumentation Engineers (SPIE) Conference Series,
  Vol. 11821, UV, X-Ray, and Gamma-Ray Space Instrumentation for Astronomy
  XXII, ed. O.~H. {Siegmund}, 118210L, \dodoi{10.1117/12.2594701}

\bibitem[{{Carmichael}(1964)}]{Carmichael64}
{Carmichael}, H. 1964, NASA Special Publication, 50, 451

\bibitem[{{Chen} {et~al.}(2015){Chen}, {Bastian}, {Shen}, {Gary}, {Krucker}, \&
  {Glesener}}]{Chen2015Particle}
{Chen}, B., {Bastian}, T.~S., {Shen}, C., {et~al.} 2015, Science, 350, 1238,
  \dodoi{10.1126/science.aac8467}

\bibitem[{{Chen} {et~al.}(2021){Chen}, {Battaglia}, {Krucker}, {Reeves}, \&
  {Glesener}}]{Chen2021Energetic}
{Chen}, B., {Battaglia}, M., {Krucker}, S., {Reeves}, K.~K., \& {Glesener}, L.
  2021, \apjl, 908, L55, \dodoi{10.3847/2041-8213/abe471}

\bibitem[{{Chen} {et~al.}(2020{\natexlab{a}}){Chen}, {Yu}, {Reeves}, \&
  {Gary}}]{Chen2020ApJ}
{Chen}, B., {Yu}, S., {Reeves}, K.~K., \& {Gary}, D.~E. 2020{\natexlab{a}},
  \apjl, 895, L50, \dodoi{10.3847/2041-8213/ab901a}

\bibitem[{{Chen} {et~al.}(2020{\natexlab{b}}){Chen}, {Shen}, {Gary}, {Reeves},
  {Fleishman}, {Yu}, {Guo}, {Krucker}, {Lin}, {Nita}, \&
  {Kong}}]{Chen2020Measurement}
{Chen}, B., {Shen}, C., {Gary}, D., {et~al.} 2020{\natexlab{b}}, in American
  Astronomical Society Meeting Abstracts, Vol. 236, American Astronomical
  Society Meeting Abstracts \#236, 112.02

\bibitem[{{Chen} {et~al.}(2024){Chen}, {Kong}, {Yu}, {Shen}, {Li}, {Guo},
  {Zhang}, {Glesener}, \& {Krucker}}]{Chen2024Energetic}
{Chen}, B., {Kong}, X., {Yu}, S., {et~al.} 2024, \apj, 971, 85,
  \dodoi{10.3847/1538-4357/ad531a}

\bibitem[{{Chen}(2016)}]{Chen2016JPlPh}
{Chen}, C.~H.~K. 2016, Journal of Plasma Physics, 82, 535820602,
  \dodoi{10.1017/S0022377816001124}

\bibitem[{{Cheng} {et~al.}(2018){Cheng}, {Li}, {Wan}, {Ding}, {Chen}, {Zhang},
  \& {Liu}}]{Cheng2018ApJ}
{Cheng}, X., {Li}, Y., {Wan}, L.~F., {et~al.} 2018, \apj, 866, 64,
  \dodoi{10.3847/1538-4357/aadd16}

\bibitem[{{Cheung} {et~al.}(2019){Cheung}, {Rempel}, {Chintzoglou}, {Chen},
  {Testa}, {Mart{\'\i}nez-Sykora}, {Sainz Dalda}, {DeRosa}, {Malanushenko},
  {Hansteen}, {De Pontieu}, {Carlsson}, {Gudiksen}, \&
  {McIntosh}}]{Cheung2019NatAs}
{Cheung}, M.~C.~M., {Rempel}, M., {Chintzoglou}, G., {et~al.} 2019, Nature
  Astronomy, 3, 160, \dodoi{10.1038/s41550-018-0629-3}

\bibitem[{{Dahlin} {et~al.}(2022){Dahlin}, {Antiochos}, {Qiu}, \&
  {DeVore}}]{Dahlin2022Variability}
{Dahlin}, J.~T., {Antiochos}, S.~K., {Qiu}, J., \& {DeVore}, C.~R. 2022, \apj,
  932, 94, \dodoi{10.3847/1538-4357/ac6e3d}

\bibitem[{{Dahlin} {et~al.}(2014){Dahlin}, {Drake}, \&
  {Swisdak}}]{Dahlin2014Mechanisms}
{Dahlin}, J.~T., {Drake}, J.~F., \& {Swisdak}, M. 2014, PhPl, 21, 092304,
  \dodoi{10.1063/1.4894484}

\bibitem[{{Dahlin} {et~al.}(2015){Dahlin}, {Drake}, \&
  {Swisdak}}]{Dahlin2015Electron}
---. 2015, PhPl, 22, 100704, \dodoi{10.1063/1.4933212}

\bibitem[{{Dahlin} {et~al.}(2017){Dahlin}, {Drake}, \&
  {Swisdak}}]{Dahlin2017Role}
---. 2017, Physics of Plasmas, 24, 092110, \dodoi{10.1063/1.4986211}

\bibitem[{{Daughton} \& {Roytershteyn}(2012)}]{Daughton2012Emerging}
{Daughton}, W., \& {Roytershteyn}, V. 2012, SSRv, 172, 271,
  \dodoi{10.1007/s11214-011-9766-z}

\bibitem[{{Daughton} {et~al.}(2011){Daughton}, {Roytershteyn}, {Karimabadi},
  {Yin}, {Albright}, {Bergen}, \& {Bowers}}]{Daughton2011Role}
{Daughton}, W., {Roytershteyn}, V., {Karimabadi}, H., {et~al.} 2011, NatPh, 7,
  539, \dodoi{10.1038/nphys1965}

\bibitem[{{Drake} {et~al.}(2019){Drake}, {Arnold}, {Swisdak}, \&
  {Dahlin}}]{Drake2019}
{Drake}, J.~F., {Arnold}, H., {Swisdak}, M., \& {Dahlin}, J.~T. 2019, Physics
  of Plasmas, 26, 012901, \dodoi{10.1063/1.5058140}

\bibitem[{{Drake} {et~al.}(2006){Drake}, {Swisdak}, {Che}, \&
  {Shay}}]{Drake2006Electron}
{Drake}, J.~F., {Swisdak}, M., {Che}, H., \& {Shay}, M.~A. 2006, Natur, 443,
  553, \dodoi{10.1038/nature05116}

\bibitem[{{Drake} {et~al.}(2013){Drake}, {Swisdak}, \&
  {Fermo}}]{Drake2013Power}
{Drake}, J.~F., {Swisdak}, M., \& {Fermo}, R. 2013, \apjl, 763, L5,
  \dodoi{10.1088/2041-8205/763/1/L5}

\bibitem[{{Drury}(1983)}]{Drury1983Introduction}
{Drury}, L.~O. 1983, Reports on Progress in Physics, 46, 973,
  \dodoi{10.1088/0034-4885/46/8/002}

\bibitem[{{Earl} {et~al.}(1988){Earl}, {Jokipii}, \&
  {Morfill}}]{Earl1988Cosmic}
{Earl}, J.~A., {Jokipii}, J.~R., \& {Morfill}, G. 1988, \apjl, 331, L91,
  \dodoi{10.1086/185242}

\bibitem[{{Egedal} {et~al.}(2013){Egedal}, {Le}, \&
  {Daughton}}]{Egedal2013Review}
{Egedal}, J., {Le}, A., \& {Daughton}, W. 2013, Physics of Plasmas, 20, 061201,
  \dodoi{10.1063/1.4811092}

\bibitem[{{Emslie} {et~al.}(2012){Emslie}, {Dennis}, {Shih}, {Chamberlin},
  {Mewaldt}, {Moore}, {Share}, {Vourlidas}, \& {Welsch}}]{Emslie2012ApJ}
{Emslie}, A.~G., {Dennis}, B.~R., {Shih}, A.~Y., {et~al.} 2012, \apj, 759, 71,
  \dodoi{10.1088/0004-637X/759/1/71}

\bibitem[{{Fleishman} {et~al.}(2020){Fleishman}, {Gary}, {Chen}, {Kuroda},
  {Yu}, \& {Nita}}]{Fleishman2020Sci}
{Fleishman}, G.~D., {Gary}, D.~E., {Chen}, B., {et~al.} 2020, Science, 367,
  278, \dodoi{10.1126/science.aax6874}

\bibitem[{{Fleishman} {et~al.}(2022){Fleishman}, {Nita}, {Chen}, {Yu}, \&
  {Gary}}]{Fleishman2022Natur}
{Fleishman}, G.~D., {Nita}, G.~M., {Chen}, B., {Yu}, S., \& {Gary}, D.~E. 2022,
  \nat, 606, 674, \dodoi{10.1038/s41586-022-04728-8}

\bibitem[{{Fletcher}(1995)}]{Fletcher1995AA}
{Fletcher}, L. 1995, \aap, 303, L9

\bibitem[{{Fletcher} \& {Hudson}(2008)}]{Fletcher2008ApJ}
{Fletcher}, L., \& {Hudson}, H.~S. 2008, \apj, 675, 1645,
  \dodoi{10.1086/527044}

\bibitem[{{Florinski} \& {Pogorelov}(2009)}]{Florinski2009Four}
{Florinski}, V., \& {Pogorelov}, N.~V. 2009, \apj, 701, 642,
  \dodoi{10.1088/0004-637X/701/1/642}

\bibitem[{{French} {et~al.}(2023){French}, {Guo}, {Zhang}, \&
  {Uzdensky}}]{French2023ApJ}
{French}, O., {Guo}, F., {Zhang}, Q., \& {Uzdensky}, D.~A. 2023, \apj, 948, 19,
  \dodoi{10.3847/1538-4357/acb7dd}

\bibitem[{{French} {et~al.}(2019){French}, {Judge}, {Matthews}, \& {van
  Driel-Gesztelyi}}]{French2019}
{French}, R.~J., {Judge}, P.~G., {Matthews}, S.~A., \& {van Driel-Gesztelyi},
  L. 2019, arXiv e-prints, arXiv:1911.12666.
\newblock \doarXiv{1911.12666}

\bibitem[{{Gan} {et~al.}(2019){Gan}, {Zhu}, {Deng}, {Li}, {Su}, {Zhang},
  {Chen}, {Zhang}, {Wu}, {Deng}, {Huang}, {Yang}, {Cui}, {Chang}, {Wang}, {Wu},
  {Yin}, {Chen}, {Fang}, {Yan}, {Lin}, {Xiong}, {Chen}, {Bao}, {Cao}, {Bai},
  {Wang}, {Chen}, {Li}, {Zhang}, {Feng}, {Su}, {Li}, {Chen}, {Li}, {Su}, {Wu},
  {Gu}, {Huang}, \& {Tang}}]{Gan2019}
{Gan}, W.-Q., {Zhu}, C., {Deng}, Y.-Y., {et~al.} 2019, Research in Astronomy
  and Astrophysics, 19, 156, \dodoi{10.1088/1674-4527/19/11/156}

\bibitem[{{Gary} {et~al.}(2023){Gary}, {Chen}, {White}, {Bastian},
  {Saint-Hilaire}, {Drake}, {Glesener}, {Yu}, {Mondal}, {Fleishman},
  {Vourlidas}, {Bale}, {Chhabra}, {Cohen}, {DeForest}, {Martinez Oliveros},
  {Ji}, {Buitrago-Casas}, {Habbal}, {Lanzerotti}, {Shaik}, {Molnar}, {Nita},
  {Emslie}, {Reardon}, {Guo}, {Oka}, {Nitta}, {Sun}, {Landi}, {Ofman}, {Lee},
  {Hudson}, {Veronig}, {Qiu}, {Leka}, {Harvey}, {Chen}, {Antiochos}, {Moore},
  {West}, {Dahlin}, {Kosovichev}, {Knipp}, {Raouafi}, {Li}, {Schad}, {Kontar},
  {Hayes}, {Shen}, {Gim{\'e}nez de Castro}, {Hannah}, {Solanki}, {Arnold},
  {Yurchyshyn}, {Cheung}, {Vourlidas}, {Martinez Pillet}, {Tarr}, {Karpen},
  {Caspi}, {Shih}, {Anan}, {Battaglia}, {Lin}, {Alaoui}, {Reeves}, {Guidoni},
  {Klimchuk}, {Kooi}, {Kazachenko}, {Tun Beltran}, {McTiernan}, {Kuroda},
  {Schonfeld}, {Kahler}, {Downs}, {Cauzzi}, {Musset}, {Gilly}, {Asai},
  {Welsch}, {Shimojo}, {Fan}, {Masuda}, {O'Donnell}, \& {Kumar}}]{FASR_WP}
{Gary}, D., {Chen}, B., {White}, S., {et~al.} 2023, in Bulletin of the American
  Astronomical Society, Vol.~55, 123, \dodoi{10.3847/25c2cfeb.7ecd0da5}

\bibitem[{{Gary}(2023)}]{Gary2023New}
{Gary}, D.~E. 2023, \araa, 61, 427, \dodoi{10.1146/annurev-astro-071221-052744}

\bibitem[{{Giacalone} \& {Jokipii}(1999)}]{Giacalone1999Transport}
{Giacalone}, J., \& {Jokipii}, J.~R. 1999, \apj, 520, 204,
  \dodoi{10.1086/307452}

\bibitem[{{Goldreich} \& {Sridhar}(1995)}]{GS95}
{Goldreich}, P., \& {Sridhar}, S. 1995, \apj, 438, 763, \dodoi{10.1086/175121}

\bibitem[{{Guidoni} {et~al.}(2016){Guidoni}, {DeVore}, {Karpen}, \&
  {Lynch}}]{Guidoni2016ApJ}
{Guidoni}, S.~E., {DeVore}, C.~R., {Karpen}, J.~T., \& {Lynch}, B.~J. 2016,
  \apj, 820, 60, \dodoi{10.3847/0004-637X/820/1/60}

\bibitem[{{Guo} \& {Giacalone}(2012)}]{Guo2012Particle}
{Guo}, F., \& {Giacalone}, J. 2012, \apj, 753, 28,
  \dodoi{10.1088/0004-637X/753/1/28}

\bibitem[{{Guo} {et~al.}(2010){Guo}, {Jokipii}, \& {Kota}}]{Guo2010}
{Guo}, F., {Jokipii}, J.~R., \& {Kota}, J. 2010, \apj, 725, 128,
  \dodoi{10.1088/0004-637X/725/1/128}

\bibitem[{{Guo} {et~al.}(2014){Guo}, {Li}, {Daughton}, \&
  {Liu}}]{Guo2014Formation}
{Guo}, F., {Li}, H., {Daughton}, W., \& {Liu}, Y.-H. 2014, PhRvL, 113, 155005

\bibitem[{{Guo} {et~al.}(2019){Guo}, {Li}, {Daughton}, {Kilian}, {Li}, {Liu},
  {Yan}, \& {Ma}}]{Guo2019Determining}
{Guo}, F., {Li}, X., {Daughton}, W., {et~al.} 2019, \apjl, 879, L23,
  \dodoi{10.3847/2041-8213/ab2a15}

\bibitem[{{Guo} {et~al.}(2021){Guo}, {Li}, {Daughton}, {Li}, {Kilian}, {Liu},
  {Zhang}, \& {Zhang}}]{Guo20213D}
---. 2021, \apj, 919, 111, \dodoi{10.3847/1538-4357/ac0918}

\bibitem[{{Guo} {et~al.}(2015){Guo}, {Liu}, {Daughton}, \&
  {Li}}]{Guo2015Particle}
{Guo}, F., {Liu}, Y.-H., {Daughton}, W., \& {Li}, H. 2015, \apj, 806, 167,
  \dodoi{10.1088/0004-637X/806/2/167}

\bibitem[{{Guo} {et~al.}(2016){Guo}, {Li}, {Li}, {Daughton}, {Zhang},
  {Lloyd-Ronning}, {Liu}, {Zhang}, \& {Deng}}]{Guo2015Efficient}
{Guo}, F., {Li}, X., {Li}, H., {et~al.} 2016, \apjl, 818, L9,
  \dodoi{10.3847/2041-8205/818/1/L9}

\bibitem[{{Hirayama}(1974)}]{Hirayama74}
{Hirayama}, T. 1974, \solphys, 34, 323, \dodoi{10.1007/BF00153671}

\bibitem[{{Holman} {et~al.}(2011){Holman}, {Aschwanden}, {Aurass}, {Battaglia},
  {Grigis}, {Kontar}, {Liu}, {Saint-Hilaire}, \& {Zharkova}}]{Holman2011}
{Holman}, G.~D., {Aschwanden}, M.~J., {Aurass}, H., {et~al.} 2011, \ssr, 159,
  107, \dodoi{10.1007/s11214-010-9680-9}

\bibitem[{{Huang} \& {Bhattacharjee}(2010)}]{Huang2010Scaling}
{Huang}, Y.-M., \& {Bhattacharjee}, A. 2010, Physics of Plasmas, 17, 062104,
  \dodoi{10.1063/1.3420208}

\bibitem[{{Huang} \& {Bhattacharjee}(2016)}]{Huang2016Turbulent}
---. 2016, \apj, 818, 20, \dodoi{10.3847/0004-637X/818/1/20}

\bibitem[{{Jeffrey}(2014)}]{Jeffrey2014PhDT}
{Jeffrey}, N. L.~S. 2014, PhD thesis, University of Glasgow

\bibitem[{{Jokipii}(1971)}]{Jokipii1971Propagation}
{Jokipii}, J.~R. 1971, Reviews of Geophysics and Space Physics, 9, 27,
  \dodoi{10.1029/RG009i001p00027}

\bibitem[{{Karpen} {et~al.}(2012){Karpen}, {Antiochos}, \&
  {DeVore}}]{Karpen2012ApJ}
{Karpen}, J.~T., {Antiochos}, S.~K., \& {DeVore}, C.~R. 2012, \apj, 760, 81,
  \dodoi{10.1088/0004-637X/760/1/81}

\bibitem[{{Kocharov} {et~al.}(2020){Kocharov}, {Pesce-Rollins}, {Laitinen},
  {Mishev}, {K{\"u}hl}, {Klassen}, {Jin}, {Omodei}, {Longo}, {Webb}, {Cane},
  {Heber}, {Vainio}, \& {Usoskin}}]{Kocharov2020ApJ}
{Kocharov}, L., {Pesce-Rollins}, M., {Laitinen}, T., {et~al.} 2020, \apj, 890,
  13, \dodoi{10.3847/1538-4357/ab684e}

\bibitem[{{Kong} {et~al.}(2020){Kong}, {Guo}, {Shen}, {Chen}, {Chen}, \&
  {Giacalone}}]{Kong2020ApJ}
{Kong}, X., {Guo}, F., {Shen}, C., {et~al.} 2020, \apjl, 905, L16,
  \dodoi{10.3847/2041-8213/abcbf5}

\bibitem[{{Kong} {et~al.}(2019){Kong}, {Guo}, {Shen}, {Chen}, {Chen}, {Musset},
  {Glesener}, {Pongkitiwanichakul}, \& {Giacalone}}]{Kong2019}
---. 2019, \apjl, 887, L37, \dodoi{10.3847/2041-8213/ab5f67}

\bibitem[{{Kong} {et~al.}(2022{\natexlab{a}}){Kong}, {Ye}, {Chen}, {Guo},
  {Shen}, {Li}, {Yu}, {Chen}, \& {Giacalone}}]{Kong2022Model}
{Kong}, X., {Ye}, J., {Chen}, B., {et~al.} 2022{\natexlab{a}}, arXiv e-prints,
  arXiv:2201.02293.
\newblock \doarXiv{2201.02293}

\bibitem[{{Kong} {et~al.}(2022{\natexlab{b}}){Kong}, {Chen}, {Guo}, {Shen},
  {Li}, {Ye}, {Zhao}, {Jiang}, {Yu}, {Chen}, \&
  {Giacalone}}]{Kong2022Numerical}
{Kong}, X., {Chen}, B., {Guo}, F., {et~al.} 2022{\natexlab{b}}, \apjl, 941,
  L22, \dodoi{10.3847/2041-8213/aca65c}

\bibitem[{{Kontar} {et~al.}(2017){Kontar}, {Perez}, {Harra}, {Kuznetsov},
  {Emslie}, {Jeffrey}, {Bian}, \& {Dennis}}]{Kontar2017}
{Kontar}, E.~P., {Perez}, J.~E., {Harra}, L.~K., {et~al.} 2017, \prl, 118,
  155101, \dodoi{10.1103/PhysRevLett.118.155101}

\bibitem[{{Kopp} \& {Pneuman}(1976)}]{Kopp76}
{Kopp}, R.~A., \& {Pneuman}, G.~W. 1976, \solphys, 50, 85,
  \dodoi{10.1007/BF00206193}

\bibitem[{{Kowal} {et~al.}(2017){Kowal}, {Falceta-Goncalves}, {Lazarian}, \&
  {Vishniac}}]{Kowal2017Statistics}
{Kowal}, G., {Falceta-Goncalves}, D.~A., {Lazarian}, A., \& {Vishniac}, E.~T.
  2017, \apj, 838, 91, \dodoi{10.3847/1538-4357/aa6001}

\bibitem[{{Krucker} {et~al.}(2010){Krucker}, {Hudson}, {Glesener}, {White},
  {Masuda}, {Wuelser}, \& {Lin}}]{Krucker2010Measure}
{Krucker}, S., {Hudson}, H.~S., {Glesener}, L., {et~al.} 2010, \apj, 714, 1108,
  \dodoi{10.1088/0004-637X/714/2/1108}

\bibitem[{{Krucker} {et~al.}(2011){Krucker}, {Hudson}, {Jeffrey}, {Battaglia},
  {Kontar}, {Benz}, {Csillaghy}, \& {Lin}}]{Krucker11High}
{Krucker}, S., {Hudson}, H.~S., {Jeffrey}, N.~L.~S., {et~al.} 2011, \apj, 739,
  96, \dodoi{10.1088/0004-637X/739/2/96}

\bibitem[{{le Roux} {et~al.}(2018){le Roux}, {Zank}, \&
  {Khabarova}}]{LeRoux2018Self}
{le Roux}, J.~A., {Zank}, G.~P., \& {Khabarova}, O.~V. 2018, \apj, 864, 158,
  \dodoi{10.3847/1538-4357/aad8b3}

\bibitem[{{le Roux} {et~al.}(2015){le Roux}, {Zank}, {Webb}, \&
  {Khabarova}}]{LeRoux2015Kinetic}
{le Roux}, J.~A., {Zank}, G.~P., {Webb}, G.~M., \& {Khabarova}, O. 2015, \apj,
  801, 112, \dodoi{10.1088/0004-637X/801/2/112}

\bibitem[{{le Roux} {et~al.}(2016){le Roux}, {Zank}, {Webb}, \&
  {Khabarova}}]{LeRoux2016Combining}
{le Roux}, J.~A., {Zank}, G.~P., {Webb}, G.~M., \& {Khabarova}, O.~V. 2016,
  \apj, 827, 47, \dodoi{10.3847/0004-637X/827/1/47}

\bibitem[{{Li} \& {Guo}(2024)}]{GPAT_GitHub}
{Li}, X., \& {Guo}, F. 2024, Global Particle Acceleration and Transport (GPAT)
  Model, \url{https://github.com/xiaocanli/stochastic-parker},  GitHub

\bibitem[{{Li} {et~al.}(2022){Li}, {Guo}, {Chen}, {Shen}, \&
  {Glesener}}]{Li2022Modeling}
{Li}, X., {Guo}, F., {Chen}, B., {Shen}, C., \& {Glesener}, L. 2022, \apj, 932,
  92, \dodoi{10.3847/1538-4357/ac6efe}

\bibitem[{{Li} {et~al.}(2018{\natexlab{a}}){Li}, {Guo}, {Li}, \&
  {Birn}}]{Li2018Roles}
{Li}, X., {Guo}, F., {Li}, H., \& {Birn}, J. 2018{\natexlab{a}}, \apj, 855, 80.
\newblock \doarXiv{1801.02255}

\bibitem[{{Li} {et~al.}(2015){Li}, {Guo}, {Li}, \& {Li}}]{Li2015Nonthermally}
{Li}, X., {Guo}, F., {Li}, H., \& {Li}, G. 2015, \apjl, 811, L24,
  \dodoi{10.1088/2041-8205/811/2/L24}

\bibitem[{{Li} {et~al.}(2017){Li}, {Guo}, {Li}, \& {Li}}]{Li2017Particle}
---. 2017, \apj, 843, 21, \dodoi{10.3847/1538-4357/aa745e}

\bibitem[{{Li} {et~al.}(2018{\natexlab{b}}){Li}, {Guo}, {Li}, \&
  {Li}}]{Li2018Large}
{Li}, X., {Guo}, F., {Li}, H., \& {Li}, S. 2018{\natexlab{b}}, \apj, 866, 4,
  \dodoi{10.3847/1538-4357/aae07b}

\bibitem[{{Li} {et~al.}(2019){Li}, {Guo}, {Li}, {Stanier}, \&
  {Kilian}}]{Li2019Formation}
{Li}, X., {Guo}, F., {Li}, H., {Stanier}, A., \& {Kilian}, P. 2019, ApJ, 884,
  118, \dodoi{10.3847/1538-4357/ab4268}

\bibitem[{{Li} {et~al.}(2021){Li}, {Guo}, \& {Liu}}]{Li2021PoP}
{Li}, X., {Guo}, F., \& {Liu}, Y.-H. 2021, Physics of Plasmas, 28, 052905,
  \dodoi{10.1063/5.0047644}

\bibitem[{{Lin} \& {Hudson}(1976)}]{Lin1976Nonthermal}
{Lin}, R.~P., \& {Hudson}, H.~S. 1976, SoPh, 50, 153,
  \dodoi{10.1007/BF00206199}

\bibitem[{{Liu} {et~al.}(2007{\natexlab{a}}){Liu}, {Lee}, {Gary}, \&
  {Wang}}]{LiuC2007}
{Liu}, C., {Lee}, J., {Gary}, D.~E., \& {Wang}, H. 2007{\natexlab{a}}, \apjl,
  658, L127, \dodoi{10.1086/513739}

\bibitem[{{Liu} {et~al.}(2007{\natexlab{b}}){Liu}, {Lee}, {Yurchyshyn}, {Deng},
  {Cho}, {Karlick{\'y}}, \& {Wang}}]{Liu07Eruption}
{Liu}, C., {Lee}, J., {Yurchyshyn}, V., {et~al.} 2007{\natexlab{b}}, \apj, 669,
  1372, \dodoi{10.1086/521644}

\bibitem[{{Loureiro} {et~al.}(2012){Loureiro}, {Samtaney}, {Schekochihin}, \&
  {Uzdensky}}]{Loureiro2012}
{Loureiro}, N.~F., {Samtaney}, R., {Schekochihin}, A.~A., \& {Uzdensky}, D.~A.
  2012, Physics of Plasmas, 19, 042303, \dodoi{10.1063/1.3703318}

\bibitem[{{McKenzie}(2000)}]{McKenzie2000}
{McKenzie}, D.~E. 2000, \solphys, 195, 381, \dodoi{10.1023/A:1005220604894}

\bibitem[{{Mondal} {et~al.}(2024){Mondal}, {Battaglia}, {Chen}, \&
  {Yu}}]{Mondal2024Joint}
{Mondal}, S., {Battaglia}, A.~F., {Chen}, B., \& {Yu}, S. 2024, \apj, 966, 208,
  \dodoi{10.3847/1538-4357/ad3910}

\bibitem[{{Montag} {et~al.}(2017){Montag}, {Egedal}, {Lichko}, \&
  {Wetherton}}]{Montag2017Impact}
{Montag}, P., {Egedal}, J., {Lichko}, E., \& {Wetherton}, B. 2017, Physics of
  Plasmas, 24, 062906, \dodoi{10.1063/1.4985302}

\bibitem[{{Omodei} {et~al.}(2018){Omodei}, {Pesce-Rollins}, {Longo},
  {Allafort}, \& {Krucker}}]{Omodei2018ApJ}
{Omodei}, N., {Pesce-Rollins}, M., {Longo}, F., {Allafort}, A., \& {Krucker},
  S. 2018, \apjl, 865, L7, \dodoi{10.3847/2041-8213/aae077}

\bibitem[{{Parker}(1965)}]{Parker1965Passage}
{Parker}, E.~N. 1965, \planss, 13, 9, \dodoi{10.1016/0032-0633(65)90131-5}

\bibitem[{{Pei} {et~al.}(2010){Pei}, {Bieber}, {Burger}, \&
  {Clem}}]{Pei2010General}
{Pei}, C., {Bieber}, J.~W., {Burger}, R.~A., \& {Clem}, J. 2010, Journal of
  Geophysical Research (Space Physics), 115, A12107,
  \dodoi{10.1029/2010JA015721}

\bibitem[{{Petrosian}(2012)}]{Petrosian2012Stochastic}
{Petrosian}, V. 2012, \ssr, 173, 535, \dodoi{10.1007/s11214-012-9900-6}

\bibitem[{{Polito} {et~al.}(2018){Polito}, {Dud{\'\i}k}, {Ka{\v{s}}parov{\'a}},
  {Dzif{\v{c}}{\'a}kov{\'a}}, {Reeves}, {Testa}, \& {Chen}}]{Polito2018ApJ}
{Polito}, V., {Dud{\'\i}k}, J., {Ka{\v{s}}parov{\'a}}, J., {et~al.} 2018, \apj,
  864, 63, \dodoi{10.3847/1538-4357/aad62d}

\bibitem[{{Qiu} {et~al.}(2023){Qiu}, {Alaoui}, {Antiochos}, {Dahlin},
  {Swisdak}, {Drake}, {Robison}, {DeVore}, \& {Uritsky}}]{Qiu2023Role}
{Qiu}, J., {Alaoui}, M., {Antiochos}, S.~K., {et~al.} 2023, \apj, 955, 34,
  \dodoi{10.3847/1538-4357/acebeb}

\bibitem[{{Reeves} {et~al.}(2017){Reeves}, {Freed}, {McKenzie}, \&
  {Savage}}]{2017ApJ...836...55R}
{Reeves}, K.~K., {Freed}, M.~S., {McKenzie}, D.~E., \& {Savage}, S.~L. 2017,
  \apj, 836, 55, \dodoi{10.3847/1538-4357/836/1/55}

\bibitem[{{Reeves} \& {Golub}(2011)}]{Reeves2011ApJ}
{Reeves}, K.~K., \& {Golub}, L. 2011, \apjl, 727, L52,
  \dodoi{10.1088/2041-8205/727/2/L52}

\bibitem[{{Reeves} {et~al.}(2019){Reeves}, {T{\"o}r{\"o}k}, {Miki{\'c}},
  {Linker}, \& {Murphy}}]{Reeves2019Exploring}
{Reeves}, K.~K., {T{\"o}r{\"o}k}, T., {Miki{\'c}}, Z., {Linker}, J., \&
  {Murphy}, N.~A. 2019, \apj, 887, 103, \dodoi{10.3847/1538-4357/ab4ce8}

\bibitem[{{Roberg-Clark} {et~al.}(2019){Roberg-Clark}, {Agapitov}, {Drake}, \&
  {Swisdak}}]{Roberg-Clark2019ApJ}
{Roberg-Clark}, G.~T., {Agapitov}, O., {Drake}, J.~F., \& {Swisdak}, M. 2019,
  \apj, 887, 190, \dodoi{10.3847/1538-4357/ab5114}

\bibitem[{{Ruan} {et~al.}(2023){Ruan}, {Yan}, \& {Keppens}}]{Ruan2023}
{Ruan}, W., {Yan}, L., \& {Keppens}, R. 2023, \apj, 947, 67,
  \dodoi{10.3847/1538-4357/ac9b4e}

\bibitem[{{Ryan} {et~al.}(2024){Ryan}, {Massa}, {Battaglia}, {Dickson}, {Su},
  {Chen}, \& {Krucker}}]{Ryan2024}
{Ryan}, D.~F., {Massa}, P., {Battaglia}, A.~F., {et~al.} 2024, \solphys, 299,
  114, \dodoi{10.1007/s11207-024-02341-8}

\bibitem[{{Samtaney} {et~al.}(2009){Samtaney}, {Loureiro}, {Uzdensky},
  {Schekochihin}, \& {Cowley}}]{Samtaney2009}
{Samtaney}, R., {Loureiro}, N.~F., {Uzdensky}, D.~A., {Schekochihin}, A.~A., \&
  {Cowley}, S.~C. 2009, \prl, 103, 105004,
  \dodoi{10.1103/PhysRevLett.103.105004}

\bibitem[{{Savage} {et~al.}(2012){Savage}, {McKenzie}, \&
  {Reeves}}]{Savage2011Re}
{Savage}, S.~L., {McKenzie}, D.~E., \& {Reeves}, K.~K. 2012, \apjl, 747, L40,
  \dodoi{10.1088/2041-8205/747/2/L40}

\bibitem[{{Seo} {et~al.}(2024){Seo}, {Guo}, {Li}, \& {Li}}]{Seo2024Proton}
{Seo}, J., {Guo}, F., {Li}, X., \& {Li}, H. 2024, \apj, 977, 146,
  \dodoi{10.3847/1538-4357/ad8e64}

\bibitem[{{Shen} {et~al.}(2022){Shen}, {Chen}, {Reeves}, {Yu}, {Polito}, \&
  {Xie}}]{Shen2022}
{Shen}, C., {Chen}, B., {Reeves}, K.~K., {et~al.} 2022, Nature Astronomy,
  \dodoi{10.1038/s41550-021-01570-2}

\bibitem[{{Shibata} {et~al.}(2023){Shibata}, {Takasao}, \&
  {Reeves}}]{Shibata2023}
{Shibata}, K., {Takasao}, S., \& {Reeves}, K.~K. 2023, \apj, 943, 106,
  \dodoi{10.3847/1538-4357/acaa9c}

\bibitem[{{Stone} {et~al.}(2008){Stone}, {Gardiner}, {Teuben}, {Hawley}, \&
  {Simon}}]{Stone2008Athena}
{Stone}, J.~M., {Gardiner}, T.~A., {Teuben}, P., {Hawley}, J.~F., \& {Simon},
  J.~B. 2008, \apjs, 178, 137, \dodoi{10.1086/588755}

\bibitem[{{Sturrock}(1966)}]{Sturrock66}
{Sturrock}, P.~A. 1966, \nat, 211, 695, \dodoi{10.1038/211695a0}

\bibitem[{{Titov} \& {D{\'e}moulin}(1999)}]{Titov99}
{Titov}, V.~S., \& {D{\'e}moulin}, P. 1999, \aap, 351, 707

\bibitem[{{Wang} {et~al.}(2025){Wang}, {Cheng}, \& {Ding}}]{Wang2025Basic}
{Wang}, Y., {Cheng}, X., \& {Ding}, M. 2025, arXiv e-prints, arXiv:2504.04648,
  \dodoi{10.48550/arXiv.2504.04648}

\bibitem[{{Wang} {et~al.}(2023){Wang}, {Cheng}, {Ding}, {Liu}, {Liu}, \&
  {Zhu}}]{Wang2023Three}
{Wang}, Y., {Cheng}, X., {Ding}, M., {et~al.} 2023, \apjl, 954, L36,
  \dodoi{10.3847/2041-8213/acf19d}

\bibitem[{{Warren} {et~al.}(2018){Warren}, {Brooks}, {Ugarte-Urra}, {Reep},
  {Crump}, \& {Doschek}}]{Warren2018ApJ}
{Warren}, H.~P., {Brooks}, D.~H., {Ugarte-Urra}, I., {et~al.} 2018, \apj, 854,
  122, \dodoi{10.3847/1538-4357/aaa9b8}

\bibitem[{{Webb} {et~al.}(2019){Webb}, {Al-Nussirat}, {Mostafavi}, {Barghouty},
  {Li}, {le Roux}, \& {Zank}}]{Webb2019}
{Webb}, G.~M., {Al-Nussirat}, S., {Mostafavi}, P., {et~al.} 2019, \apj, 881,
  123, \dodoi{10.3847/1538-4357/ab2fca}

\bibitem[{{Webb} {et~al.}(2018){Webb}, {Barghouty}, {Hu}, \& {le
  Roux}}]{Webb2018}
{Webb}, G.~M., {Barghouty}, A.~F., {Hu}, Q., \& {le Roux}, J.~A. 2018, \apj,
  855, 31, \dodoi{10.3847/1538-4357/aaae6c}

\bibitem[{{Xie} {et~al.}(2024){Xie}, {Li}, {Reeves}, \&
  {Gou}}]{2024FrASS..1183746X}
{Xie}, X., {Li}, G., {Reeves}, K.~K., \& {Gou}, T. 2024, Frontiers in Astronomy
  and Space Sciences, 11, 1383746, \dodoi{10.3389/fspas.2024.1383746}

\bibitem[{{Xie} \& {Reeves}(2023)}]{Xie2023Heating}
{Xie}, X., \& {Reeves}, K.~K. 2023, \apj, 942, 28,
  \dodoi{10.3847/1538-4357/ac9f47}

\bibitem[{{Xie} {et~al.}(2022){Xie}, {Reeves}, {Shen}, \&
  {Ingram}}]{Xie2022ApJ}
{Xie}, X., {Reeves}, K.~K., {Shen}, C., \& {Ingram}, J.~D. 2022, \apj, 933, 15,
  \dodoi{10.3847/1538-4357/ac695d}

\bibitem[{{Xie} {et~al.}(2025){Xie}, {Shen}, {Reeves}, {Chen}, {Li}, {Guo},
  {Yu}, {Wei}, \& {Dong}}]{Xie2025ApJ}
{Xie}, X., {Shen}, C., {Reeves}, K.~K., {et~al.} 2025, \apjl, 984, L27,
  \dodoi{10.3847/2041-8213/adc91b}

\bibitem[{{Yin} {et~al.}(2024){Yin}, {Drake}, \&
  {Swisdak}}]{Yin2024Simultaneous}
{Yin}, Z., {Drake}, J.~F., \& {Swisdak}, M. 2024, \apj, 974, 74,
  \dodoi{10.3847/1538-4357/ad7131}

\bibitem[{{Yu} \& {Chen}(2019)}]{Yu2019Possible}
{Yu}, S., \& {Chen}, B. 2019, \apj, 872, 71, \dodoi{10.3847/1538-4357/aaff6d}

\bibitem[{{Yu} {et~al.}(2020){Yu}, {Chen}, {Reeves}, {Gary}, {Musset},
  {Fleishman}, {Nita}, \& {Glesener}}]{Yu2020ApJ}
{Yu}, S., {Chen}, B., {Reeves}, K.~K., {et~al.} 2020, \apj, 900, 17,
  \dodoi{10.3847/1538-4357/aba8a6}

\bibitem[{{Zank} {et~al.}(2014){Zank}, {le Roux}, {Webb}, {Dosch}, \&
  {Khabarova}}]{Zank2014Particle}
{Zank}, G.~P., {le Roux}, J.~A., {Webb}, G.~M., {Dosch}, A., \& {Khabarova}, O.
  2014, \apj, 797, 28, \dodoi{10.1088/0004-637X/797/1/28}

\bibitem[{{Zank} {et~al.}(2015){Zank}, {Hunana}, {Mostafavi}, {Le Roux}, {Li},
  {Webb}, {Khabarova}, {Cummings}, {Stone}, \& {Decker}}]{Zank2015Diffusive}
{Zank}, G.~P., {Hunana}, P., {Mostafavi}, P., {et~al.} 2015, \apj, 814, 137,
  \dodoi{10.1088/0004-637X/814/2/137}

\bibitem[{{Zhang}(1999)}]{Zhang1999Markov}
{Zhang}, M. 1999, \apj, 513, 409, \dodoi{10.1086/306857}

\bibitem[{{Zhang} {et~al.}(2024{\natexlab{a}}){Zhang}, {Guo}, {Daughton}, {Li},
  {Le}, {Phan}, \& {Desai}}]{Zhang2024PhRvL}
{Zhang}, Q., {Guo}, F., {Daughton}, W., {et~al.} 2024{\natexlab{a}}, \prl, 132,
  115201, \dodoi{10.1103/PhysRevLett.132.115201}

\bibitem[{{Zhang} {et~al.}(2021){Zhang}, {Guo}, {Daughton}, {Li}, \&
  {Li}}]{Zhang2021Efficient}
{Zhang}, Q., {Guo}, F., {Daughton}, W., {Li}, H., \& {Li}, X. 2021, \prl, 127,
  185101, \dodoi{10.1103/PhysRevLett.127.185101}

\bibitem[{{Zhang} {et~al.}(2024{\natexlab{b}}){Zhang}, {Guo}, {Daughton}, {Li},
  \& {Li}}]{Zhang2024Plasma}
{Zhang}, Q., {Guo}, F., {Daughton}, W., {Li}, X., \& {Li}, H.
  2024{\natexlab{b}}, \apj, 974, 47, \dodoi{10.3847/1538-4357/ad6561}

\bibitem[{{Zhang} {et~al.}(2019){Zhang}, {Chen}, {Wu}, {Chang}, {Hu}, {Su},
  {Zhang}, {Wang}, {Liang}, {Ma}, {Guo}, {Cai}, {Zhang}, {Huang}, {Peng},
  {Tang}, {Zhao}, {Zhou}, {Wang}, {Song}, {Ma}, {Xu}, {Yang}, {Lu}, {He},
  {Tao}, {Ma}, {Lv}, {Bai}, {Cao}, {Huang}, \& {Gan}}]{ZhangZ2019}
{Zhang}, Z., {Chen}, D.-Y., {Wu}, J., {et~al.} 2019, Research in Astronomy and
  Astrophysics, 19, 160, \dodoi{10.1088/1674-4527/19/11/160}

\bibitem[{{Zhao} {et~al.}(2019){Zhao}, {Zank}, {Chen}, {Hu}, {le Roux}, {Du},
  \& {Adhikari}}]{Zhao2019Particle}
{Zhao}, L.-L., {Zank}, G.~P., {Chen}, Y., {et~al.} 2019, \apj, 872, 4,
  \dodoi{10.3847/1538-4357/aafcb2}

\bibitem[{{Zhao} {et~al.}(2018){Zhao}, {Zank}, {Khabarova}, {Du}, {Chen},
  {Adhikari}, \& {Hu}}]{Zhao2018Unusual}
{Zhao}, L.-L., {Zank}, G.~P., {Khabarova}, O., {et~al.} 2018, \apjl, 864, L34,
  \dodoi{10.3847/2041-8213/aaddf6}

\end{thebibliography}

\end{document}